\documentclass[runningheads]{llncs}

\usepackage[T1]{fontenc}
\usepackage{graphicx}
\usepackage[draft]{todonotes}
\usepackage{amscd}
\usepackage{amsfonts}
\usepackage{amsmath}
\usepackage{amssymb}
\usepackage{booktabs}
\usepackage{comment}
\usepackage{environ}
\usepackage{graphicx}
\usepackage{hyperref}
\usepackage[capitalise]{cleveref}
\usepackage{latexsym}
\usepackage{marginnote}
\usepackage{multirow}
\usepackage{paralist}
\usepackage{pgfplots}
\usepackage{pifont}
\usepackage{placeins}
\usepackage{rotate} 
\usepackage[FIGBOTCAP,TABBOTCAP,center,nooneline]{subfigure}
\usepackage{tikz}
\usepackage{times} 
\usepackage{url}
\usepackage{varwidth}
\usepackage{verbatim} 
\usepackage{wrapfig}
\usepackage{xcolor,colortbl}
\usepackage{xspace}
\usetikzlibrary{intersections, backgrounds, calc}
\usepackage{color}

\usepackage{cancel}

\definecolor {snow}                {rgb}{1.00,0.98,0.98}
\definecolor {ghostwhite}          {rgb}{0.97,0.97,1.00}
\definecolor {whitesmoke}          {rgb}{0.96,0.96,0.96}
\definecolor {gainsboro}           {rgb}{0.86,0.86,0.86}
\definecolor {floralwhite}         {rgb}{1.00,0.98,0.94}
\definecolor {oldlace}             {rgb}{0.99,0.96,0.90}
\definecolor {linen}               {rgb}{0.98,0.94,0.90}
\definecolor {antiquewhite}        {rgb}{0.98,0.92,0.84}
\definecolor {papayawhip}          {rgb}{1.00,0.94,0.84}
\definecolor {blanchedalmond}      {rgb}{1.00,0.92,0.80}
\definecolor {bisque}              {rgb}{1.00,0.89,0.77}
\definecolor {peachpuff}           {rgb}{1.00,0.85,0.73}
\definecolor {navajowhite}         {rgb}{1.00,0.87,0.68}
\definecolor {moccasin}            {rgb}{1.00,0.89,0.71}
\definecolor {cornsilk}            {rgb}{1.00,0.97,0.86}
\definecolor {ivory}               {rgb}{1.00,1.00,0.94}
\definecolor {lemonchiffon}        {rgb}{1.00,0.98,0.80}
\definecolor {seashell}            {rgb}{1.00,0.96,0.93}
\definecolor {honeydew}            {rgb}{0.94,1.00,0.94}
\definecolor {mintcream}           {rgb}{0.96,1.00,0.98}
\definecolor {azure}               {rgb}{0.94,1.00,1.00}
\definecolor {aliceblue}           {rgb}{0.94,0.97,1.00}
\definecolor {lavender}            {rgb}{0.90,0.90,0.98}
\definecolor {lavenderblush}       {rgb}{1.00,0.94,0.96}
\definecolor {mistyrose}           {rgb}{1.00,0.89,0.88}
\definecolor {white}               {rgb}{1.00,1.00,1.00}
\definecolor {black}               {rgb}{0.00,0.00,0.00}
\definecolor {darkslategray}       {rgb}{0.18,0.31,0.31}
\definecolor {dimgray}             {rgb}{0.41,0.41,0.41}
\definecolor {slategray}           {rgb}{0.44,0.50,0.56}
\definecolor {lightslategray}      {rgb}{0.47,0.53,0.60}
\definecolor {gray}                {rgb}{0.75,0.75,0.75}
\definecolor {lightgrey}           {rgb}{0.83,0.83,0.83}
\definecolor {midnightblue}        {rgb}{0.10,0.10,0.44}
\definecolor {navy}                {rgb}{0.00,0.00,0.50}
\definecolor {cornflowerblue}      {rgb}{0.39,0.58,0.93}
\definecolor {darkslateblue}       {rgb}{0.28,0.24,0.55}
\definecolor {slateblue}           {rgb}{0.42,0.35,0.80}
\definecolor {mediumslateblue}     {rgb}{0.48,0.41,0.93}
\definecolor {lightslateblue}      {rgb}{0.52,0.44,1.00}
\definecolor {mediumblue}          {rgb}{0.00,0.00,0.80}
\definecolor {royalblue}           {rgb}{0.25,0.41,0.88}
\definecolor {blue}                {rgb}{0.00,0.00,1.00}
\definecolor {dodgerblue}          {rgb}{0.12,0.56,1.00}
\definecolor {deepskyblue}         {rgb}{0.00,0.75,1.00}
\definecolor {skyblue}             {rgb}{0.53,0.81,0.92}
\definecolor {lightskyblue}        {rgb}{0.53,0.81,0.98}
\definecolor {steelblue}           {rgb}{0.27,0.51,0.71}
\definecolor {lightsteelblue}      {rgb}{0.69,0.77,0.87}
\definecolor {lightblue}           {rgb}{0.68,0.85,0.90}
\definecolor {powderblue}          {rgb}{0.69,0.88,0.90}
\definecolor {paleturquoise}       {rgb}{0.69,0.93,0.93}
\definecolor {darkturquoise}       {rgb}{0.00,0.81,0.82}
\definecolor {mediumturquoise}     {rgb}{0.28,0.82,0.80}
\definecolor {turquoise}           {rgb}{0.25,0.88,0.82}
\definecolor {cyan}                {rgb}{0.00,1.00,1.00}
\definecolor {lightcyan}           {rgb}{0.88,1.00,1.00}
\definecolor {cadetblue}           {rgb}{0.37,0.62,0.63}
\definecolor {mediumaquamarine}    {rgb}{0.40,0.80,0.67}
\definecolor {aquamarine}          {rgb}{0.50,1.00,0.83}
\definecolor {darkgreen}           {rgb}{0.00,0.39,0.00}
\definecolor {darkolivegreen}      {rgb}{0.33,0.42,0.18}
\definecolor {darkseagreen}        {rgb}{0.56,0.74,0.56}
\definecolor {seagreen}            {rgb}{0.18,0.55,0.34}
\definecolor {mediumseagreen}      {rgb}{0.24,0.70,0.44}
\definecolor {lightseagreen}       {rgb}{0.13,0.70,0.67}
\definecolor {palegreen}           {rgb}{0.60,0.98,0.60}
\definecolor {springgreen}         {rgb}{0.00,1.00,0.50}
\definecolor {lawngreen}           {rgb}{0.49,0.99,0.00}
\definecolor {green}               {rgb}{0.00,1.00,0.00}
\definecolor {chartreuse}          {rgb}{0.50,1.00,0.00}
\definecolor {mediumspringgreen}   {rgb}{0.00,0.98,0.60}
\definecolor {greenyellow}         {rgb}{0.68,1.00,0.18}
\definecolor {limegreen}           {rgb}{0.20,0.80,0.20}
\definecolor {yellowgreen}         {rgb}{0.60,0.80,0.20}
\definecolor {forestgreen}         {rgb}{0.13,0.55,0.13}
\definecolor {olivedrab}           {rgb}{0.42,0.56,0.14}
\definecolor {darkkhaki}           {rgb}{0.74,0.72,0.42}
\definecolor {khaki}               {rgb}{0.94,0.90,0.55}
\definecolor {palegoldenrod}       {rgb}{0.93,0.91,0.67}
\definecolor {lightgoldenrodyellow} {rgb}{0.98,0.98,0.82}
\definecolor {lightyellow}         {rgb}{1.00,1.00,0.88}
\definecolor {yellow}              {rgb}{1.00,1.00,0.00}
\definecolor {gold}                {rgb}{1.00,0.84,0.00}
\definecolor {lightgoldenrod}      {rgb}{0.93,0.87,0.51}
\definecolor {goldenrod}           {rgb}{0.85,0.65,0.13}
\definecolor {darkgoldenrod}       {rgb}{0.72,0.53,0.04}
\definecolor {rosybrown}           {rgb}{0.74,0.56,0.56}
\definecolor {indianred}           {rgb}{0.80,0.36,0.36}
\definecolor {saddlebrown}         {rgb}{0.55,0.27,0.07}
\definecolor {sienna}              {rgb}{0.63,0.32,0.18}
\definecolor {peru}                {rgb}{0.80,0.52,0.25}
\definecolor {burlywood}           {rgb}{0.87,0.72,0.53}
\definecolor {beige}               {rgb}{0.96,0.96,0.86}
\definecolor {wheat}               {rgb}{0.96,0.87,0.70}
\definecolor {sandybrown}          {rgb}{0.96,0.64,0.38}
\definecolor {tan}                 {rgb}{0.82,0.71,0.55}
\definecolor {chocolate}           {rgb}{0.82,0.41,0.12}
\definecolor {firebrick}           {rgb}{0.70,0.13,0.13}
\definecolor {brown}               {rgb}{0.65,0.16,0.16}
\definecolor {darksalmon}          {rgb}{0.91,0.59,0.48}
\definecolor {salmon}              {rgb}{0.98,0.50,0.45}
\definecolor {lightsalmon}         {rgb}{1.00,0.63,0.48}
\definecolor {orange}              {rgb}{1.00,0.65,0.00}
\definecolor {darkorange}          {rgb}{1.00,0.55,0.00}
\definecolor {coral}               {rgb}{1.00,0.50,0.31}
\definecolor {lightcoral}          {rgb}{0.94,0.50,0.50}
\definecolor {tomato}              {rgb}{1.00,0.39,0.28}
\definecolor {orangered}           {rgb}{1.00,0.27,0.00}
\definecolor {red}                 {rgb}{1.00,0.00,0.00}
\definecolor {hotpink}             {rgb}{1.00,0.41,0.71}
\definecolor {deeppink}            {rgb}{1.00,0.08,0.58}
\definecolor {pink}                {rgb}{1.00,0.75,0.80}
\definecolor {lightpink}           {rgb}{1.00,0.71,0.76}
\definecolor {palevioletred}       {rgb}{0.86,0.44,0.58}
\definecolor {maroon}              {rgb}{0.69,0.19,0.38}
\definecolor {mediumvioletred}     {rgb}{0.78,0.08,0.52}
\definecolor {violetred}           {rgb}{0.82,0.13,0.56}
\definecolor {magenta}             {rgb}{1.00,0.00,1.00}
\definecolor {violet}              {rgb}{0.93,0.51,0.93}
\definecolor {plum}                {rgb}{0.87,0.63,0.87}
\definecolor {orchid}              {rgb}{0.85,0.44,0.84}
\definecolor {mediumorchid}        {rgb}{0.73,0.33,0.83}
\definecolor {darkorchid}          {rgb}{0.60,0.20,0.80}
\definecolor {darkviolet}          {rgb}{0.58,0.00,0.83}
\definecolor {blueviolet}          {rgb}{0.54,0.17,0.89}
\definecolor {purple}              {rgb}{0.63,0.13,0.94}
\definecolor {mediumpurple}        {rgb}{0.58,0.44,0.86}
\definecolor {thistle}             {rgb}{0.85,0.75,0.85}
\definecolor {snow2}               {rgb}{0.93,0.91,0.91}
\definecolor {snow3}               {rgb}{0.80,0.79,0.79}
\definecolor {snow4}               {rgb}{0.55,0.54,0.54}
\definecolor {seashell2}           {rgb}{0.93,0.90,0.87}
\definecolor {seashell3}           {rgb}{0.80,0.77,0.75}
\definecolor {seashell4}           {rgb}{0.55,0.53,0.51}
\definecolor {antiquewhite1}       {rgb}{1.00,0.94,0.86}
\definecolor {antiquewhite2}       {rgb}{0.93,0.87,0.80}
\definecolor {antiquewhite3}       {rgb}{0.80,0.75,0.69}
\definecolor {antiquewhite4}       {rgb}{0.55,0.51,0.47}
\definecolor {bisque2}             {rgb}{0.93,0.84,0.72}
\definecolor {bisque3}             {rgb}{0.80,0.72,0.62}
\definecolor {bisque4}             {rgb}{0.55,0.49,0.42}
\definecolor {peachpuff2}          {rgb}{0.93,0.80,0.68}
\definecolor {peachpuff3}          {rgb}{0.80,0.69,0.58}
\definecolor {peachpuff4}          {rgb}{0.55,0.47,0.40}
\definecolor {navajowhite2}        {rgb}{0.93,0.81,0.63}
\definecolor {navajowhite3}        {rgb}{0.80,0.70,0.55}
\definecolor {navajowhite4}        {rgb}{0.55,0.47,0.37}
\definecolor {lemonchiffon2}       {rgb}{0.93,0.91,0.75}
\definecolor {lemonchiffon3}       {rgb}{0.80,0.79,0.65}
\definecolor {lemonchiffon4}       {rgb}{0.55,0.54,0.44}
\definecolor {cornsilk2}           {rgb}{0.93,0.91,0.80}
\definecolor {cornsilk3}           {rgb}{0.80,0.78,0.69}
\definecolor {cornsilk4}           {rgb}{0.55,0.53,0.47}
\definecolor {ivory2}              {rgb}{0.93,0.93,0.88}
\definecolor {ivory3}              {rgb}{0.80,0.80,0.76}
\definecolor {ivory4}              {rgb}{0.55,0.55,0.51}
\definecolor {honeydew2}           {rgb}{0.88,0.93,0.88}
\definecolor {honeydew3}           {rgb}{0.76,0.80,0.76}
\definecolor {honeydew4}           {rgb}{0.51,0.55,0.51}
\definecolor {lavenderblush2}      {rgb}{0.93,0.88,0.90}
\definecolor {lavenderblush3}      {rgb}{0.80,0.76,0.77}
\definecolor {lavenderblush4}      {rgb}{0.55,0.51,0.53}
\definecolor {mistyrose2}          {rgb}{0.93,0.84,0.82}
\definecolor {mistyrose3}          {rgb}{0.80,0.72,0.71}
\definecolor {mistyrose4}          {rgb}{0.55,0.49,0.48}
\definecolor {azure2}              {rgb}{0.88,0.93,0.93}
\definecolor {azure3}              {rgb}{0.76,0.80,0.80}
\definecolor {azure4}              {rgb}{0.51,0.55,0.55}
\definecolor {slateblue1}          {rgb}{0.51,0.44,1.00}
\definecolor {slateblue2}          {rgb}{0.48,0.40,0.93}
\definecolor {slateblue3}          {rgb}{0.41,0.35,0.80}
\definecolor {slateblue4}          {rgb}{0.28,0.24,0.55}
\definecolor {royalblue1}          {rgb}{0.28,0.46,1.00}
\definecolor {royalblue2}          {rgb}{0.26,0.43,0.93}
\definecolor {royalblue3}          {rgb}{0.23,0.37,0.80}
\definecolor {royalblue4}          {rgb}{0.15,0.25,0.55}
\definecolor {blue2}               {rgb}{0.00,0.00,0.93}
\definecolor {blue4}               {rgb}{0.00,0.00,0.55}
\definecolor {dodgerblue2}         {rgb}{0.11,0.53,0.93}
\definecolor {dodgerblue3}         {rgb}{0.09,0.45,0.80}
\definecolor {dodgerblue4}         {rgb}{0.06,0.31,0.55}
\definecolor {steelblue1}          {rgb}{0.39,0.72,1.00}
\definecolor {steelblue2}          {rgb}{0.36,0.67,0.93}
\definecolor {steelblue3}          {rgb}{0.31,0.58,0.80}
\definecolor {steelblue4}          {rgb}{0.21,0.39,0.55}
\definecolor {deepskyblue2}        {rgb}{0.00,0.70,0.93}
\definecolor {deepskyblue3}        {rgb}{0.00,0.60,0.80}
\definecolor {deepskyblue4}        {rgb}{0.00,0.41,0.55}
\definecolor {skyblue1}            {rgb}{0.53,0.81,1.00}
\definecolor {skyblue2}            {rgb}{0.49,0.75,0.93}
\definecolor {skyblue3}            {rgb}{0.42,0.65,0.80}
\definecolor {skyblue4}            {rgb}{0.29,0.44,0.55}
\definecolor {lightskyblue1}       {rgb}{0.69,0.89,1.00}
\definecolor {lightskyblue2}       {rgb}{0.64,0.83,0.93}
\definecolor {lightskyblue3}       {rgb}{0.55,0.71,0.80}
\definecolor {lightskyblue4}       {rgb}{0.38,0.48,0.55}
\definecolor {slategray1}          {rgb}{0.78,0.89,1.00}
\definecolor {slategray2}          {rgb}{0.73,0.83,0.93}
\definecolor {slategray3}          {rgb}{0.62,0.71,0.80}
\definecolor {slategray4}          {rgb}{0.42,0.48,0.55}
\definecolor {lightsteelblue1}     {rgb}{0.79,0.88,1.00}
\definecolor {lightsteelblue2}     {rgb}{0.74,0.82,0.93}
\definecolor {lightsteelblue3}     {rgb}{0.64,0.71,0.80}
\definecolor {lightsteelblue4}     {rgb}{0.43,0.48,0.55}
\definecolor {lightblue1}          {rgb}{0.75,0.94,1.00}
\definecolor {lightblue2}          {rgb}{0.70,0.87,0.93}
\definecolor {lightblue3}          {rgb}{0.60,0.75,0.80}
\definecolor {lightblue4}          {rgb}{0.41,0.51,0.55}
\definecolor {lightcyan2}          {rgb}{0.82,0.93,0.93}
\definecolor {lightcyan3}          {rgb}{0.71,0.80,0.80}
\definecolor {lightcyan4}          {rgb}{0.48,0.55,0.55}
\definecolor {paleturquoise1}      {rgb}{0.73,1.00,1.00}
\definecolor {paleturquoise2}      {rgb}{0.68,0.93,0.93}
\definecolor {paleturquoise3}      {rgb}{0.59,0.80,0.80}
\definecolor {paleturquoise4}      {rgb}{0.40,0.55,0.55}
\definecolor {cadetblue1}          {rgb}{0.60,0.96,1.00}
\definecolor {cadetblue2}          {rgb}{0.56,0.90,0.93}
\definecolor {cadetblue3}          {rgb}{0.48,0.77,0.80}
\definecolor {cadetblue4}          {rgb}{0.33,0.53,0.55}
\definecolor {turquoise1}          {rgb}{0.00,0.96,1.00}
\definecolor {turquoise2}          {rgb}{0.00,0.90,0.93}
\definecolor {turquoise3}          {rgb}{0.00,0.77,0.80}
\definecolor {turquoise4}          {rgb}{0.00,0.53,0.55}
\definecolor {cyan2}               {rgb}{0.00,0.93,0.93}
\definecolor {cyan3}               {rgb}{0.00,0.80,0.80}
\definecolor {cyan4}               {rgb}{0.00,0.55,0.55}
\definecolor {darkslategray1}      {rgb}{0.59,1.00,1.00}
\definecolor {darkslategray2}      {rgb}{0.55,0.93,0.93}
\definecolor {darkslategray3}      {rgb}{0.47,0.80,0.80}
\definecolor {darkslategray4}      {rgb}{0.32,0.55,0.55}
\definecolor {aquamarine2}         {rgb}{0.46,0.93,0.78}
\definecolor {aquamarine4}         {rgb}{0.27,0.55,0.45}
\definecolor {darkseagreen1}       {rgb}{0.76,1.00,0.76}
\definecolor {darkseagreen2}       {rgb}{0.71,0.93,0.71}
\definecolor {darkseagreen3}       {rgb}{0.61,0.80,0.61}
\definecolor {darkseagreen4}       {rgb}{0.41,0.55,0.41}
\definecolor {seagreen1}           {rgb}{0.33,1.00,0.62}
\definecolor {seagreen2}           {rgb}{0.31,0.93,0.58}
\definecolor {seagreen3}           {rgb}{0.26,0.80,0.50}
\definecolor {palegreen1}          {rgb}{0.60,1.00,0.60}
\definecolor {palegreen2}          {rgb}{0.56,0.93,0.56}
\definecolor {palegreen3}          {rgb}{0.49,0.80,0.49}
\definecolor {palegreen4}          {rgb}{0.33,0.55,0.33}
\definecolor {springgreen2}        {rgb}{0.00,0.93,0.46}
\definecolor {springgreen3}        {rgb}{0.00,0.80,0.40}
\definecolor {springgreen4}        {rgb}{0.00,0.55,0.27}
\definecolor {green2}              {rgb}{0.00,0.93,0.00}
\definecolor {green3}              {rgb}{0.00,0.80,0.00}
\definecolor {green4}              {rgb}{0.00,0.55,0.00}
\definecolor {chartreuse2}         {rgb}{0.46,0.93,0.00}
\definecolor {chartreuse3}         {rgb}{0.40,0.80,0.00}
\definecolor {chartreuse4}         {rgb}{0.27,0.55,0.00}
\definecolor {olivedrab1}          {rgb}{0.75,1.00,0.24}
\definecolor {olivedrab2}          {rgb}{0.70,0.93,0.23}
\definecolor {olivedrab4}          {rgb}{0.41,0.55,0.13}
\definecolor {darkolivegreen1}     {rgb}{0.79,1.00,0.44}
\definecolor {darkolivegreen2}     {rgb}{0.74,0.93,0.41}
\definecolor {darkolivegreen3}     {rgb}{0.64,0.80,0.35}
\definecolor {darkolivegreen4}     {rgb}{0.43,0.55,0.24}
\definecolor {khaki1}              {rgb}{1.00,0.96,0.56}
\definecolor {khaki2}              {rgb}{0.93,0.90,0.52}
\definecolor {khaki3}              {rgb}{0.80,0.78,0.45}
\definecolor {khaki4}              {rgb}{0.55,0.53,0.31}
\definecolor {lightgoldenrod1}     {rgb}{1.00,0.93,0.55}
\definecolor {lightgoldenrod2}     {rgb}{0.93,0.86,0.51}
\definecolor {lightgoldenrod3}     {rgb}{0.80,0.75,0.44}
\definecolor {lightgoldenrod4}     {rgb}{0.55,0.51,0.30}
\definecolor {lightyellow2}        {rgb}{0.93,0.93,0.82}
\definecolor {lightyellow3}        {rgb}{0.80,0.80,0.71}
\definecolor {lightyellow4}        {rgb}{0.55,0.55,0.48}
\definecolor {yellow2}             {rgb}{0.93,0.93,0.00}
\definecolor {yellow3}             {rgb}{0.80,0.80,0.00}
\definecolor {yellow4}             {rgb}{0.55,0.55,0.00}
\definecolor {gold2}               {rgb}{0.93,0.79,0.00}
\definecolor {gold3}               {rgb}{0.80,0.68,0.00}
\definecolor {gold4}               {rgb}{0.55,0.46,0.00}
\definecolor {goldenrod1}          {rgb}{1.00,0.76,0.15}
\definecolor {goldenrod2}          {rgb}{0.93,0.71,0.13}
\definecolor {goldenrod3}          {rgb}{0.80,0.61,0.11}
\definecolor {goldenrod4}          {rgb}{0.55,0.41,0.08}
\definecolor {darkgoldenrod1}      {rgb}{1.00,0.73,0.06}
\definecolor {darkgoldenrod2}      {rgb}{0.93,0.68,0.05}
\definecolor {darkgoldenrod3}      {rgb}{0.80,0.58,0.05}
\definecolor {darkgoldenrod4}      {rgb}{0.55,0.40,0.03}
\definecolor {rosybrown1}          {rgb}{1.00,0.76,0.76}
\definecolor {rosybrown2}          {rgb}{0.93,0.71,0.71}
\definecolor {rosybrown3}          {rgb}{0.80,0.61,0.61}
\definecolor {rosybrown4}          {rgb}{0.55,0.41,0.41}
\definecolor {indianred1}          {rgb}{1.00,0.42,0.42}
\definecolor {indianred2}          {rgb}{0.93,0.39,0.39}
\definecolor {indianred3}          {rgb}{0.80,0.33,0.33}
\definecolor {indianred4}          {rgb}{0.55,0.23,0.23}
\definecolor {sienna1}             {rgb}{1.00,0.51,0.28}
\definecolor {sienna2}             {rgb}{0.93,0.47,0.26}
\definecolor {sienna3}             {rgb}{0.80,0.41,0.22}
\definecolor {sienna4}             {rgb}{0.55,0.28,0.15}
\definecolor {burlywood1}          {rgb}{1.00,0.83,0.61}
\definecolor {burlywood2}          {rgb}{0.93,0.77,0.57}
\definecolor {burlywood3}          {rgb}{0.80,0.67,0.49}
\definecolor {burlywood4}          {rgb}{0.55,0.45,0.33}
\definecolor {wheat1}              {rgb}{1.00,0.91,0.73}
\definecolor {wheat2}              {rgb}{0.93,0.85,0.68}
\definecolor {wheat3}              {rgb}{0.80,0.73,0.59}
\definecolor {wheat4}              {rgb}{0.55,0.49,0.40}
\definecolor {tan1}                {rgb}{1.00,0.65,0.31}
\definecolor {tan2}                {rgb}{0.93,0.60,0.29}
\definecolor {tan4}                {rgb}{0.55,0.35,0.17}
\definecolor {chocolate1}          {rgb}{1.00,0.50,0.14}
\definecolor {chocolate2}          {rgb}{0.93,0.46,0.13}
\definecolor {chocolate3}          {rgb}{0.80,0.40,0.11}
\definecolor {firebrick1}          {rgb}{1.00,0.19,0.19}
\definecolor {firebrick2}          {rgb}{0.93,0.17,0.17}
\definecolor {firebrick3}          {rgb}{0.80,0.15,0.15}
\definecolor {firebrick4}          {rgb}{0.55,0.10,0.10}
\definecolor {brown1}              {rgb}{1.00,0.25,0.25}
\definecolor {brown2}              {rgb}{0.93,0.23,0.23}
\definecolor {brown3}              {rgb}{0.80,0.20,0.20}
\definecolor {brown4}              {rgb}{0.55,0.14,0.14}
\definecolor {salmon1}             {rgb}{1.00,0.55,0.41}
\definecolor {salmon2}             {rgb}{0.93,0.51,0.38}
\definecolor {salmon3}             {rgb}{0.80,0.44,0.33}
\definecolor {salmon4}             {rgb}{0.55,0.30,0.22}
\definecolor {lightsalmon2}        {rgb}{0.93,0.58,0.45}
\definecolor {lightsalmon3}        {rgb}{0.80,0.51,0.38}
\definecolor {lightsalmon4}        {rgb}{0.55,0.34,0.26}
\definecolor {orange2}             {rgb}{0.93,0.60,0.00}
\definecolor {orange3}             {rgb}{0.80,0.52,0.00}
\definecolor {orange4}             {rgb}{0.55,0.35,0.00}
\definecolor {darkorange1}         {rgb}{1.00,0.50,0.00}
\definecolor {darkorange2}         {rgb}{0.93,0.46,0.00}
\definecolor {darkorange3}         {rgb}{0.80,0.40,0.00}
\definecolor {darkorange4}         {rgb}{0.55,0.27,0.00}
\definecolor {coral1}              {rgb}{1.00,0.45,0.34}
\definecolor {coral2}              {rgb}{0.93,0.42,0.31}
\definecolor {coral3}              {rgb}{0.80,0.36,0.27}
\definecolor {coral4}              {rgb}{0.55,0.24,0.18}
\definecolor {tomato2}             {rgb}{0.93,0.36,0.26}
\definecolor {tomato3}             {rgb}{0.80,0.31,0.22}
\definecolor {tomato4}             {rgb}{0.55,0.21,0.15}
\definecolor {orangered2}          {rgb}{0.93,0.25,0.00}
\definecolor {orangered3}          {rgb}{0.80,0.22,0.00}
\definecolor {orangered4}          {rgb}{0.55,0.15,0.00}
\definecolor {red2}                {rgb}{0.93,0.00,0.00}
\definecolor {red3}                {rgb}{0.80,0.00,0.00}
\definecolor {red4}                {rgb}{0.55,0.00,0.00}
\definecolor {deeppink2}           {rgb}{0.93,0.07,0.54}
\definecolor {deeppink3}           {rgb}{0.80,0.06,0.46}
\definecolor {deeppink4}           {rgb}{0.55,0.04,0.31}
\definecolor {hotpink1}            {rgb}{1.00,0.43,0.71}
\definecolor {hotpink2}            {rgb}{0.93,0.42,0.65}
\definecolor {hotpink3}            {rgb}{0.80,0.38,0.56}
\definecolor {hotpink4}            {rgb}{0.55,0.23,0.38}
\definecolor {pink1}               {rgb}{1.00,0.71,0.77}
\definecolor {pink2}               {rgb}{0.93,0.66,0.72}
\definecolor {pink3}               {rgb}{0.80,0.57,0.62}
\definecolor {pink4}               {rgb}{0.55,0.39,0.42}
\definecolor {lightpink1}          {rgb}{1.00,0.68,0.73}
\definecolor {lightpink2}          {rgb}{0.93,0.64,0.68}
\definecolor {lightpink3}          {rgb}{0.80,0.55,0.58}
\definecolor {lightpink4}          {rgb}{0.55,0.37,0.40}
\definecolor {palevioletred1}      {rgb}{1.00,0.51,0.67}
\definecolor {palevioletred2}      {rgb}{0.93,0.47,0.62}
\definecolor {palevioletred3}      {rgb}{0.80,0.41,0.54}
\definecolor {palevioletred4}      {rgb}{0.55,0.28,0.36}
\definecolor {maroon1}             {rgb}{1.00,0.20,0.70}
\definecolor {maroon2}             {rgb}{0.93,0.19,0.65}
\definecolor {maroon3}             {rgb}{0.80,0.16,0.56}
\definecolor {maroon4}             {rgb}{0.55,0.11,0.38}
\definecolor {violetred1}          {rgb}{1.00,0.24,0.59}
\definecolor {violetred2}          {rgb}{0.93,0.23,0.55}
\definecolor {violetred3}          {rgb}{0.80,0.20,0.47}
\definecolor {violetred4}          {rgb}{0.55,0.13,0.32}
\definecolor {magenta2}            {rgb}{0.93,0.00,0.93}
\definecolor {magenta3}            {rgb}{0.80,0.00,0.80}
\definecolor {magenta4}            {rgb}{0.55,0.00,0.55}
\definecolor {orchid1}             {rgb}{1.00,0.51,0.98}
\definecolor {orchid2}             {rgb}{0.93,0.48,0.91}
\definecolor {orchid3}             {rgb}{0.80,0.41,0.79}
\definecolor {orchid4}             {rgb}{0.55,0.28,0.54}
\definecolor {plum1}               {rgb}{1.00,0.73,1.00}
\definecolor {plum2}               {rgb}{0.93,0.68,0.93}
\definecolor {plum3}               {rgb}{0.80,0.59,0.80}
\definecolor {plum4}               {rgb}{0.55,0.40,0.55}
\definecolor {mediumorchid1}       {rgb}{0.88,0.40,1.00}
\definecolor {mediumorchid2}       {rgb}{0.82,0.37,0.93}
\definecolor {mediumorchid3}       {rgb}{0.71,0.32,0.80}
\definecolor {mediumorchid4}       {rgb}{0.48,0.22,0.55}
\definecolor {darkorchid1}         {rgb}{0.75,0.24,1.00}
\definecolor {darkorchid2}         {rgb}{0.70,0.23,0.93}
\definecolor {darkorchid3}         {rgb}{0.60,0.20,0.80}
\definecolor {darkorchid4}         {rgb}{0.41,0.13,0.55}
\definecolor {purple1}             {rgb}{0.61,0.19,1.00}
\definecolor {purple2}             {rgb}{0.57,0.17,0.93}
\definecolor {purple3}             {rgb}{0.49,0.15,0.80}
\definecolor {purple4}             {rgb}{0.33,0.10,0.55}
\definecolor {mediumpurple1}       {rgb}{0.67,0.51,1.00}
\definecolor {mediumpurple2}       {rgb}{0.62,0.47,0.93}
\definecolor {mediumpurple3}       {rgb}{0.54,0.41,0.80}
\definecolor {mediumpurple4}       {rgb}{0.36,0.28,0.55}
\definecolor {thistle1}            {rgb}{1.00,0.88,1.00}
\definecolor {thistle2}            {rgb}{0.93,0.82,0.93}
\definecolor {thistle3}            {rgb}{0.80,0.71,0.80}
\definecolor {thistle4}            {rgb}{0.55,0.48,0.55}
\definecolor {gray1}               {rgb}{0.01,0.01,0.01}
\definecolor {gray2}               {rgb}{0.02,0.02,0.02}
\definecolor {gray3}               {rgb}{0.03,0.03,0.03}
\definecolor {gray4}               {rgb}{0.04,0.04,0.04}
\definecolor {gray5}               {rgb}{0.05,0.05,0.05}
\definecolor {gray6}               {rgb}{0.06,0.06,0.06}
\definecolor {gray7}               {rgb}{0.07,0.07,0.07}
\definecolor {gray8}               {rgb}{0.08,0.08,0.08}
\definecolor {gray9}               {rgb}{0.09,0.09,0.09}
\definecolor {gray10}              {rgb}{0.10,0.10,0.10}
\definecolor {gray11}              {rgb}{0.11,0.11,0.11}
\definecolor {gray12}              {rgb}{0.12,0.12,0.12}
\definecolor {gray13}              {rgb}{0.13,0.13,0.13}
\definecolor {gray14}              {rgb}{0.14,0.14,0.14}
\definecolor {gray15}              {rgb}{0.15,0.15,0.15}
\definecolor {gray16}              {rgb}{0.16,0.16,0.16}
\definecolor {gray17}              {rgb}{0.17,0.17,0.17}
\definecolor {gray18}              {rgb}{0.18,0.18,0.18}
\definecolor {gray19}              {rgb}{0.19,0.19,0.19}
\definecolor {gray20}              {rgb}{0.20,0.20,0.20}
\definecolor {gray21}              {rgb}{0.21,0.21,0.21}
\definecolor {gray22}              {rgb}{0.22,0.22,0.22}
\definecolor {gray23}              {rgb}{0.23,0.23,0.23}
\definecolor {gray24}              {rgb}{0.24,0.24,0.24}
\definecolor {gray25}              {rgb}{0.25,0.25,0.25}
\definecolor {gray26}              {rgb}{0.26,0.26,0.26}
\definecolor {gray27}              {rgb}{0.27,0.27,0.27}
\definecolor {gray28}              {rgb}{0.28,0.28,0.28}
\definecolor {gray29}              {rgb}{0.29,0.29,0.29}
\definecolor {gray30}              {rgb}{0.30,0.30,0.30}
\definecolor {gray31}              {rgb}{0.31,0.31,0.31}
\definecolor {gray32}              {rgb}{0.32,0.32,0.32}
\definecolor {gray33}              {rgb}{0.33,0.33,0.33}
\definecolor {gray34}              {rgb}{0.34,0.34,0.34}
\definecolor {gray35}              {rgb}{0.35,0.35,0.35}
\definecolor {gray36}              {rgb}{0.36,0.36,0.36}
\definecolor {gray37}              {rgb}{0.37,0.37,0.37}
\definecolor {gray38}              {rgb}{0.38,0.38,0.38}
\definecolor {gray39}              {rgb}{0.39,0.39,0.39}
\definecolor {gray40}              {rgb}{0.40,0.40,0.40}
\definecolor {gray42}              {rgb}{0.42,0.42,0.42}
\definecolor {gray43}              {rgb}{0.43,0.43,0.43}
\definecolor {gray44}              {rgb}{0.44,0.44,0.44}
\definecolor {gray45}              {rgb}{0.45,0.45,0.45}
\definecolor {gray46}              {rgb}{0.46,0.46,0.46}
\definecolor {gray47}              {rgb}{0.47,0.47,0.47}
\definecolor {gray48}              {rgb}{0.48,0.48,0.48}
\definecolor {gray49}              {rgb}{0.49,0.49,0.49}
\definecolor {gray50}              {rgb}{0.50,0.50,0.50}
\definecolor {gray51}              {rgb}{0.51,0.51,0.51}
\definecolor {gray52}              {rgb}{0.52,0.52,0.52}
\definecolor {gray53}              {rgb}{0.53,0.53,0.53}
\definecolor {gray54}              {rgb}{0.54,0.54,0.54}
\definecolor {gray55}              {rgb}{0.55,0.55,0.55}
\definecolor {gray56}              {rgb}{0.56,0.56,0.56}
\definecolor {gray57}              {rgb}{0.57,0.57,0.57}
\definecolor {gray58}              {rgb}{0.58,0.58,0.58}
\definecolor {gray59}              {rgb}{0.59,0.59,0.59}
\definecolor {gray60}              {rgb}{0.60,0.60,0.60}
\definecolor {gray61}              {rgb}{0.61,0.61,0.61}
\definecolor {gray62}              {rgb}{0.62,0.62,0.62}
\definecolor {gray63}              {rgb}{0.63,0.63,0.63}
\definecolor {gray64}              {rgb}{0.64,0.64,0.64}
\definecolor {gray65}              {rgb}{0.65,0.65,0.65}
\definecolor {gray66}              {rgb}{0.66,0.66,0.66}
\definecolor {gray67}              {rgb}{0.67,0.67,0.67}
\definecolor {gray68}              {rgb}{0.68,0.68,0.68}
\definecolor {gray69}              {rgb}{0.69,0.69,0.69}
\definecolor {gray70}              {rgb}{0.70,0.70,0.70}
\definecolor {gray71}              {rgb}{0.71,0.71,0.71}
\definecolor {gray72}              {rgb}{0.72,0.72,0.72}
\definecolor {gray73}              {rgb}{0.73,0.73,0.73}
\definecolor {gray74}              {rgb}{0.74,0.74,0.74}
\definecolor {gray75}              {rgb}{0.75,0.75,0.75}
\definecolor {gray76}              {rgb}{0.76,0.76,0.76}
\definecolor {gray77}              {rgb}{0.77,0.77,0.77}
\definecolor {gray78}              {rgb}{0.78,0.78,0.78}
\definecolor {gray79}              {rgb}{0.79,0.79,0.79}
\definecolor {gray80}              {rgb}{0.80,0.80,0.80}
\definecolor {gray81}              {rgb}{0.81,0.81,0.81}
\definecolor {gray82}              {rgb}{0.82,0.82,0.82}
\definecolor {gray83}              {rgb}{0.83,0.83,0.83}
\definecolor {gray84}              {rgb}{0.84,0.84,0.84}
\definecolor {gray85}              {rgb}{0.85,0.85,0.85}
\definecolor {gray86}              {rgb}{0.86,0.86,0.86}
\definecolor {gray87}              {rgb}{0.87,0.87,0.87}
\definecolor {gray88}              {rgb}{0.88,0.88,0.88}
\definecolor {gray89}              {rgb}{0.89,0.89,0.89}
\definecolor {gray90}              {rgb}{0.90,0.90,0.90}
\definecolor {gray91}              {rgb}{0.91,0.91,0.91}
\definecolor {gray92}              {rgb}{0.92,0.92,0.92}
\definecolor {gray93}              {rgb}{0.93,0.93,0.93}
\definecolor {gray94}              {rgb}{0.94,0.94,0.94}
\definecolor {gray95}              {rgb}{0.95,0.95,0.95}
\definecolor {gray97}              {rgb}{0.97,0.97,0.97}
\definecolor {gray98}              {rgb}{0.98,0.98,0.98}
\definecolor {gray99}              {rgb}{0.99,0.99,0.99}
\definecolor {darkgrey}            {rgb}{0.66,0.66,0.66}

\newcommand{\new}[1]{{\blue #1}\/}

\newcommand{\resp}[1]{[resp. #1]}

\newcommand{\TODO}[1]{{}}

\newcommand{\ignore}[1]{}

\newcommand{\RSTODO}[1]{{\bf \textcolor{darkgreen}{{\fbox{RS TODO:} #1}}}}
\renewcommand{\RSTODO}[1]{}

 \newcommand{\ignoreinshort}[1]{}
 \newcommand{\ignoreinlong}[1]{{#1}}

\def\makenewenumerate#1#2{%
\newcounter{cnt#1}
\newenvironment{#1}%
{\begin{list}{\makebox[0pt][r]{#2}}%
{\setlength{\itemsep}{0pt}%
 \setlength{\parsep}{.2em}%
 \setlength{\leftmargin}{1.5em}%
 \setlength{\labelwidth}{.4em}%
 \usecounter{cnt#1}}}
{\end{list}}}

\makenewenumerate{myenumerate}{\arabic{cntmyenumerate}.}

\makenewenumerate{renumerate}{\rm(\roman{cntrenumerate})}
\makenewenumerate{renumerateprime}{\rm(\roman{cntrenumerateprime}$'$)}
\makenewenumerate{renumeratesecond}{\rm(\roman{cntrenumeratesecond}$''$)}

\makenewenumerate{aenumerate}{({\it\alph{cntaenumerate}})}
\makenewenumerate{aenumerateprime}{({\it\alph{cntaenumerateprime}$'$})}
\makenewenumerate{aenumeratesecond}{({\it\alph{cntaenumeratesecond}$''$})}

\def\newplaintheorem#1#2{%
\newtheorem{#1plain}{#2}[section]%
\newenvironment{#1}{\begin{#1plain}\rm }{\end{#1plain}}}

\newtheorem{Property}{Property}

\newplaintheorem{notation}{Notation}
\newplaintheorem{cexample}{Counter-Example}

\newcommand{\sref}[1]{\S{}\ref{#1}}
\newcommand{\noi}{\noindent}

\newcommand{\set}[1]{\ensuremath{\{{#1}\}}\xspace}
\newcommand{\imp}{\ensuremath{\rightarrow}\xspace}
\newcommand{\limp}{\ensuremath{\leftarrow}\xspace}
\renewcommand{\iff}{\ensuremath{\leftrightarrow}\xspace}
\newcommand{\defas}{\ensuremath{\stackrel{\text{\tiny def}}{=}}\xspace}
\newcommand{\thus}{\ensuremath{\Longrightarrow}\xspace}

\newcommand{\pos}{\phantom{\neg}}

\newcommand{\trueval}{{\ensuremath{\mathsf{true}}}}
\newcommand{\falseval}{{\ensuremath{\mathsf{false}}}}

\newcommand\mysout{\bgroup \markoverwith{{-}}\ULon}
\newcommand\nosout{\bgroup \markoverwith{{ }}\ULon}
\definecolor{mygray}{rgb}{0.90,0.90,0.90}
\definecolor{mywhite}{rgb}{1.00,1.00,1.00}

\newcommand{\vi}{\ensuremath{\varphi}\xspace}

\newcommand{\T}{\ensuremath{\mathcal{T}}\xspace}

\newcommand{\mathsat}{\textsc{MathSAT}\xspace}

\renewcommand{\new}[1]{{\em #1}}
\renewcommand{\RSTODO}[1]{{\bf \textcolor{darkgreen}{{\fbox{RS TODO:} #1}}}}
\renewcommand{\paragraph}[1]{\smallskip\noindent{\emph{#1}}}

\newcommand{\alla}{\ensuremath{{\mathbf{A}}}\xspace}
\newcommand{\allb}{\ensuremath{{\mathbf{B}}}\xspace}
\renewcommand{\vi}{\ensuremath{\varphi_{[\alla]}}\xspace}
\newcommand{\vione}{\ensuremath{\varphi'_{[\alla]}}\xspace}
\newcommand{\vitwo}{\ensuremath{\varphi''_{[\alla]}}\xspace}
\renewcommand{\vi}{\ensuremath{\varphi}\xspace}
\renewcommand{\vione}{\ensuremath{\varphi_{1}}\xspace}
\renewcommand{\vitwo}{\ensuremath{\varphi_{2}}\xspace}
\newcommand{\ps}{\ensuremath{\psi_{[\alla,\allb]}}\xspace}

\renewcommand{\trueval}{\ensuremath{\mbox{{\sf T}}\xspace}}
\renewcommand{\falseval}{\ensuremath{\mbox{{\sf F}}\xspace}}
\newcommand{\unknown}{\ensuremath{\mbox{{\sf ?}}\xspace}}

\newcommand{\tval}{\trueval}
\newcommand{\fval}{\falseval}
\newcommand{\uval}{\unknown}

\newcommand{\andmu}{\ensuremath{\bigwedge\!\mu}\xspace}
\newcommand{\andeta}{\ensuremath{\bigwedge\!\eta}\xspace}

\newcommand{\apply}[2]{\ensuremath{#1|_{#2}}\xspace}
\newcommand{\applymuvi}{\apply{\vi}{\mu}\xspace}
\newcommand{\applyetavi}{\apply{\vi}{\eta}\xspace}

\newcommand{\muof}[1]{\ensuremath{\mu(#1)}}

\newcommand{\oksym}{{\Large \ensuremath{\textcolor{darkgreen}{\checkmark}}}\xspace}
\newcommand{\noksym}{{\LARGE \ensuremath{\textcolor{red}{\mathbf{\times}}}}\xspace}

\newcommand{\shannon}[2]{\ensuremath{#1|_{\exists#2}}\xspace}
\renewcommand{\shannon}[2]{\ensuremath{{\sf SE}[{\exists#2}.#1}]\xspace}
\newcommand{\shannonpsib}{\shannon{\psi}{\allb}}

\newcommand{\predabsphi}{\ensuremath{{\sf PredAbs}(\phi,\set{\phi_i}_{i=1}^N)}}
\renewcommand{\predabsphi}{\ensuremath{{\sf PredAbs}(\phi,{\mathbf \Phi})}}

\newcommand{\genericentails}{\mid\!\simeq}
\newcommand{\entails}{\ensuremath{\models}}
\newcommand{\verifies}{\vdash}
\renewcommand{\verifies}{\Vdash}
\renewcommand{\verifies}{\ensuremath{\mid\!\approx}}

\newcommand{\verbverify}{validate}
\newcommand{\verbverified}{validated}
\newcommand{\verbverifies}{validates}
\newcommand{\verbverifying}{validating}
\newcommand{\verbverification}{verification}
\newcommand{\Verbverification}{Verification}

\renewcommand{\verbverify}{evaluate to true}
\renewcommand{\verbverified}{evaluated to true}
\renewcommand{\verbverifies}{evaluates to true}
\renewcommand{\verbverifying}{evaluating to true}
\renewcommand{\verbverification}{evaluation to true}
\renewcommand{\Verbverification}{Evaluation to true}

\renewcommand{\verbverify}{verify}
\renewcommand{\verbverified}{verified}
\renewcommand{\verbverifies}{verifies}
\renewcommand{\verbverifying}{verifying}
\renewcommand{\verbverification}{verification}
\renewcommand{\Verbverification}{Verification}

\renewcommand{\RSTODO}[1]{\todo[size=\tiny,color=green!40]{{\small{#1}}}}

\newcommand{\cnfnamed}[1]{\ensuremath{\mathsf{CNF_{#1}}}}
\newcommand{\DeMorganCNF}{\cnfnamed{DM}}
\newcommand{\TseitinCNF}{\cnfnamed{Ts}}
\newcommand{\PlaistedCNF}{\cnfnamed{PG}}
\newcommand{\NNF}[1]{\ensuremath{\mathsf{NNF}(#1)}}
\newcommand{\NNFna}[1]{\ensuremath{\mathsf{NNF}}}

\newcommand{\mainwffone}{\ensuremath{(A_1\wedge  A_2)\vee(A_1\wedge \neg A_2)}}
\newcommand{\mainwfftwo}{\ensuremath{(\neg A_3\wedge  A_4)\vee(\neg A_3\wedge \neg A_4)}}
\newcommand{\mainwff}{\ensuremath{(\mainwffone)\wedge(\mainwfftwo)}}

\newcommand{\wffl}{\ensuremath{(A_1\wedge  A_2)}}
\newcommand{\wffr}{\ensuremath{(A_1\wedge \neg A_2)}}
\newcommand{\wffltwo}{\ensuremath{(\neg A_3\wedge  A_4)}}
\newcommand{\wffrtwo}{\ensuremath{(\neg A_3\wedge \neg A_4)}} 
\newcommand{\Aone}{\ensuremath{A_1}}
\newcommand{\nAone}{\ensuremath{\neg A_1}}
\newcommand{\Atwo}{\ensuremath{A_2}}
\newcommand{\nAtwo}{\ensuremath{\neg A_2}}  
\newcommand{\Athree}{\ensuremath{A_3}}
\newcommand{\nAthree}{\ensuremath{\neg A_3}}
\newcommand{\Afour}{\ensuremath{A_4}}
\newcommand{\nAfour}{\ensuremath{\neg A_4}} 
\newcommand{\singlemu}{\set{A_1,\neg A_3}}
\newcommand{\singlemuset}{\set{\set{A_1,\neg A_3}}}
\newcommand{\allmus}{
  \set{
    \set{A_1,A_2,\neg A_3,A_4},
    \set{A_1,A_2,\neg A_3,\neg A_4},
    \set{A_1,\neg A_2,\neg A_3,A_4},
    \set{A_1,\neg A_2,\neg A_3,\neg A_4}}}

\newcommand{\longversion}{true}
\ifthenelse{\equal{\longversion}{true}}
{%
  \renewcommand{\ignoreinshort}[1]{\textcolor{midnightblue}{#1}}
  \renewcommand{\ignoreinlong}[1]{}
  \specialcomment{IGNOREINSHORT}{\begingroup\color{midnightblue}}{\endgroup}
  \excludecomment{IGNOREINLONG}
}%
{%
 \renewcommand{\ignoreinshort}[1]{}
 \renewcommand{\ignoreinlong}[1]{\textcolor{midnightblue}{#1}}
 \excludecomment{IGNOREINSHORT}
  \specialcomment{IGNOREINLONG}{\begingroup\color{black}}{\endgroup}
}
 \excludecomment{IGNORE}

\excludecomment{TOBEDROPPED}

\renewcommand{\TODO}[1]{\todo[inline,color=green!40]{{\small{#1}}}}

\begin{document}


\pagestyle{plain}
\pagenumbering{arabic}

\title{%
  Entailment vs. \Verbverification{} 
\\for   Partial-assignment Satisfiability and Enumeration
}

\author{
Roberto Sebastiani~\orcidID{0000-0002-0989-6101}
}
\authorrunning{R. Sebastiani}

\institute{%
DISI, University of Trento, Italy.\\    
\email{roberto.sebastiani@unitn.it}
}

\maketitle
 \begin{abstract}
Many procedures for 
SAT-related problems, in particular for those requiring the complete
enumeration of satisfying truth assignments,
rely their efficiency and effectiveness
on the detection of (possibly small) {\em partial} assignments satisfying an input
formula. 
Surprisingly,
{there seems to be no unique universally-agreed definition of formula satisfaction by
a partial assignment in the literature}.

In this paper, we analyze in depth the issue of satisfaction by partial assignments,
raising a flag about
some ambiguities and subtleties of this concept, and investigating their
practical consequences.
We identify two alternative notions that are implicitly used in the literature,
namely {\em verification} and {\em entailment},
which coincide if applied to CNF formulas, but differ and
present complementary properties if
applied to non-CNF or to existentially-quantified formulas. 
We show that, although the former is easier to check and as such is implicitly used
by most current search procedures, the latter has better theoretical
properties, and
can improve the efficiency and effectiveness of enumeration
procedures.

 \end{abstract}

\section{Introduction}
\label{sec:intro}
\emph{Motivations.}
Many search procedures for SAT-related problems (e.g., Analytic
Tableaux \cite{smullyan1}, DPLL \cite{davis7}, circuit AllSAT~\cite{friedAllSATCombinationalCircuits2023}) and many formula compilers
(e.g., d-DNNFs
\cite{darwicheCompilerDeterministicDecomposable2002,darwicheKnowledgeCompilationMap2002},
OBDDs \cite{bryant2} and SDDs \cite{darwicheSDDNewCanonical2011})
rely their efficiency and effectiveness on the detection of {\em partial} truth
assignments $\mu$ satisfying an input propositional formula $\vi$,
which allows to state that not only 
$\vi$ is satisfiable, but also
all total
assignments extending $\mu$ satisfy \vi.
In particular,
when it comes to SAT-based problems requiring
the {\em complete enumeration} of satisfying assignments (e.g. \#SAT
\cite{GomesSS21}, Lazy SMT \cite{sebastiani07,BarrettSST21}, OMT \cite{sebastiani15_optimathsat}, AllSAT and AllSMT
\cite{lahiriSMTTechniquesFast2006,cavada_fmcad07_predabs,friedAllSATCombinationalCircuits2023},
\#SMT \cite{PhanM15,chisticov15}, Projected \#SAT
\cite{azizExistsSATProjected2015}, Projected AllSAT/AllSMT
\cite{lahiriSMTTechniquesFast2006,gebserSolutionEnumerationProjected2009,spallittaDisjointPartialEnumeration2024},
knowledge compilation into d-DNNF{} \cite{darwicheCompilerDeterministicDecomposable2002,lagniezLeveragingDecisionDNNFCompilation2024},
satisfiability in modal and description logics
\cite{SebastianiT21}, Weighted Model Integration
\cite{morettin_aij19,spallittaEnhancingSMTbasedWeighted2024a}),
the ability of enumerating satisfying {partial} assignments which are
as small as possible is essential, because
each of them avoids the enumeration of the whole subtree of total
assignments extending it, whose size is exponential in the number of
unassigned atoms.

We start by raising 
a
very-basic
question: \emph{What should we mean by ``a partial
assignment $\mu$ satisfies a formula $\vi$''?} 
We notice that, quite surprisingly and despite its
widespread implicit usage in algorithms, 
{there is no unique and universally-agreed definition for
formula satisfaction by partial assignments
in the literature}:
most authors do not define it 
  explicitly;
a few others define it only when dealing with
CNF formulas (e.g.  \cite{BuningK21});
the very few who define it, 
  adopt one of the following distinct 
  definitions: 
\begin{itemize}
\item 
[``\emph{applying $\mu$ to $\vi$ makes $\vi$ true}''] (e.g.,
  \cite{DarwicheDNNF-ACM01,friedAllSATCombinationalCircuits2023}).
We say that \emph{$\mu$ \verbverifies{} $\vi$};~\footnote{%
The etymology of ``to verify'' derives from Latin ``{\em verum facere}'', meaning ``to make true''.} 
\item
   [``\emph{all total assignments extending $\mu$ satisfy $\vi$}'']
   (e.g., \cite{gs-infocomp2000,sebastiani07}).
   We say that \emph{$\mu$ entails $\vi$}.
\end{itemize}
%
Notice that this is not simply an issue of
the meaning of the word ``satisfy'': regardless which verb we
may  use for it (e.g. ``satisfy'', ``entail'',  ``verify'',
``imply'',...), it should be desirable to have a unique and universally-agreed criterion
to establish it.
\ignore{%
  \footnote{In 2020 I ran an online poll among SAT and SMT PC members about the meaning of partial-assignment satisfiability: 35\% said entailment, 35\% said verification, 30\% said both (!).}}

\paragraph{Contributions.}
In this paper, we 
analyze in depth
the notion of {\em partial-assignment
satisfiability}, in particular when dealing with  non-CNF and
existentially-quantified formulas,
raising  a flag about the ambiguities and subtleties
of this concept, and
investigating their consequences.
Our contributions are both theoretical and practical.

\paragraph{Theoretical analysis.}
We show, 
analyze and discuss
the following theoretical facts.


\begin{aenumerate}
\item Whereas for (tautology-free\footnote{A CNF formula is
tautology-free if it contains no tautological clause in the form
$(l\vee \neg l\vee ...)$. Since tautological clauses can be easily
removed upfront by simple preprocessing, hereafter we often assume
w.l.o.g. that CNF formulas are tautology-free.})
CNF formulas
partial-assignment satisfiability can be indifferently be interpreted
either as \verbverification{} or as entailment
because in this case the two concepts coincide,
{\em for non-CNF formulas \verbverification{} is
strictly stronger than entailment}, and they have complementary
properties.

\item
Whereas two equivalent 
  formulas are always entailed by the
same partial assignments,
{\em they are not always
  \verbverified{} by the same partial assignments}.

\item \Verbverification{} checks that the residual $\applymuvi$ of a
  formula \vi{} under a partial assignment $\mu$
  ---i.e., the formula obtained by  applying the assigned truth values from $\mu$ to the atoms in $\vi$---
  is \emph{the true formula}, whereas
  entailment checks it is \emph{a valid formula}. Thus
  \verbverification{} can be computationally much cheaper to check than
  entailment. 

\item
$(a)$-$(c)$ apply also to
existentially-quantified formulas, \emph{even those in CNF}.

\item
$(a)$-$(c)$ apply also to
existentially-quantified formulas resulting from CNF-ization.

\ignore{  
CNF-izing upfront the formulas 
\cite{tseitin1,plaistedStructurepreservingClauseForm1986} does not
fix 
issues $(a)$-$(c)$, since fresh atoms are introduced, and
\emph{a partial assignment over the original atoms
  may
  entail the existentially-quantified 
  CNF-ized formula without   \verbverifying{} it.}
} 

\end{aenumerate}
We stress the fact that $(b)$ (and $(d)$ and $(e)$ consequently) would be an embarrassing fact
if we adopted \verbverification{} as the definition of  partial-assignment
satisfiability. As such, we champion the idea
that the latter should be defined as
entailment \cite{gs-infocomp2000,sebastiani07,SebastianiT21},
and that \verbverification{} should be considered an easy-to-check sufficient 
condition for it. 



\paragraph{Practical consequences.}
The above facts have
the following practical consequences.

\begin{enumerate}
\item In  satisfiability, we need
  to find only one total assignment $\eta$ extending $\mu$, so that
  entailment produces no benefits
  wrt. \verbverification{} and is more expensive $(c)$. 
  Also, due to $(a)$, when the input problem is natively in CNF, the 
  distinction between \verbverification{} and entailment is not relevant, because these two concepts
  coincide.~%
\ignore{\footnote{We conjecture that this is the main reason why the distinction between
\verbverification{} and entailment has been long overlooked in the literature.}}

\end{enumerate}

\noindent
Consequently, both the algorithms for satisfiability and those for enumeration with CNF
formulas typically use \verbverification{} as satisfiability criterion for
the current partial assignments, because it
is much cheaper and easier to implement than entailment.~%
\ignore{\footnote{For satisfiability of CNF formulas, CDCL SAT solvers produce
  {\em total} truth
assignments.}}
This is the case, e.g., of classical procedures like  Analytic
Tableaux \cite{smullyan1} and DPLL \cite{davis7,armando5}, 
or 
enumeration, counting, or knowledge compilation procedures, e.g. 
\cite{darwicheCompilerDeterministicDecomposable2002,lahiriSMTTechniquesFast2006,gebserSolutionEnumerationProjected2009,azizExistsSATProjected2015,friedAllSATCombinationalCircuits2023,lagniezLeveragingDecisionDNNFCompilation2024}.
A few notable exceptions are the {\sf Dualiza} procedure
\cite{mohleDualizingProjectedModel2018} and the procedures we
described in 
\cite{moehle20,moehle21,friedEntailingGeneralizationBoosts2024};
also OBDDs \cite{bryant2} and SDDs \cite{darwicheSDDNewCanonical2011}
formula compilers implicitly use entailment to prune branches so that
to guarantee canonicity (see below and 
\sref{sec:practical}).

The scenario changes when we deal with
enumeration-based algorithms applied to non-CNF formulas, or to
existentially-quantified formulas, or to CNF-ized formulas.
\begin{enumerate}
  \setcounter{enumi}{1}
\item Due to $(a)$, a partial assignment may entail a formula without
  \verbverifying{} it. Thus, adopting entailment as partial-assignment satisfiability
  criterion during the search allows for detecting satisfiability earlier than with
  verification, 
  and thus for producing smaller partial truth assignments.
\end{enumerate}
\noindent
Although this fact is not very interesting for  satisfiability,
it may become fundamental for enumeration, because the detection of a
satisfying partial assignment avoids the enumeration of the whole subtree of total assignments
extending it, whose size is exponential in the number of unassigned
atoms: \emph{the earlier satisfiability is detected, the (up to
exponentially) fewer assignments are enumerated}.


\begin{enumerate}
  \setcounter{enumi}{2}
\item Due to $(d)$, a partial assignment may entail an
  existentially-quantified formula, even a CNF one, without
  \verbverifying{} it. Thus, as with non-CNF formulas $(b)$, adopting entailment as partial-assignment satisfiability
  criterion during the search allows for detecting satisfiability earlier than with
  verification.
\end{enumerate}

\noindent
This is important because in many application domains
fundamental operations ---like {\em  
  pre-image computation} in symbolic model checking
(see e.g. \cite{burch1}) or {\em predicate abstraction} in SW verification
(see e.g. \cite{graf_predabs97,lahiriSMTTechniquesFast2006}) or {\em projected
  enumeration} (see e.g. \cite{lahiriSMTTechniquesFast2006,gebserSolutionEnumerationProjected2009,spallittaDisjointPartialEnumeration2024})
or {\em projected model counting} (see e.g. \cite{azizExistsSATProjected2015}) ---
require dealing with existentially-quantified formulas and with
the enumeration of partial assignments satisfying them.

\begin{enumerate}
  \setcounter{enumi}{3}
\item Due to $(e)$, CNF-izing upfront the non-CNF formula with the standard
  techniques \cite{tseitin1,plaistedStructurepreservingClauseForm1986}
  does not fix issues $(a)$-$(c)$, since fresh atoms are
  introduced, and \emph{a partial assignment over the original
    atoms may entail the existentially-quantified CNF-ized
    formula without \verbverifying{} it.}
\end{enumerate}

\noindent
This is important because it shows that,
although verification and entailment coincide for CNF formulas, 
Fact
2. above cannot be fixed by simply CNF-izing a formula
upfront and running a CNF enumeration procedure based on partial-assignment
verification, projecting out the fresh variables introduced by the CNF-ization.

\begin{enumerate}
  \setcounter{enumi}{4}
\item Due to $(c)$, checking \verbverification{}
is polynomial, whereas checking
entailment is co-NP-complete, since it consists of checking the validity of
the residual formula. 
\end{enumerate}

\noindent
This is the main argument in favour of \verbverification{}
vs. entailment. 
We notice, however, that if $\mu$ entails $\vi$, then the residual of
$\vi$ under $\mu$ 
is typically much smaller than \vi with a much
smaller search space, since its
atoms are only (a subset of) those that are not assigned
by $\mu$. Notice also that, when this happens, entailment prevents
from enumerating a number of assignments which is exponential in the
number of the atoms in the residual. Therefore, unlike with
 satisfiability, for enumeration the tradeoff may be favourable
to entailment checks.


Based on the considerations above, not only we suggest to adopt
entailment 
 rather than \verbverification{} as unique definition of partial-assignment
 satisfiability, but also we champion its usage inside
 enumeration-based search
 procedures. 
 \ignore{
To this extent, in the last years we have proposed techniques for efficiently implementing
 entailment within enumeration
 procedures~\cite{moehle20,moehle21,friedEntailingGeneralizationBoosts2024},
 which are mostly based on \cite{mohleDualizingProjectedModel2018};
 in particular, in \cite{friedEntailingGeneralizationBoosts2024} we
 have enhanced the enumeration tool for circuits from
 \cite{friedAllSATCombinationalCircuits2023} by substituting
 \verbverification{} checks with entailment checks, obtaining drastic
 improvements in terms of both time efficiency and size of 
 partial-assignment sets .
}

  
\paragraph{Related Work.}
\label{sec:related}
This paper is a revised version of a 2020  unpublished paper
\cite{sebastianiAreYouSatisfied2020}, which at that time was not published due to
lack of algorithmic support and empirical evidence.
These came afterwards. Based on the theoretical analysis in 
\cite{sebastianiAreYouSatisfied2020} in combination with the idea of dualized search
from \cite{FazekasSB16,mohleDualizingProjectedModel2018},
in~\cite{moehle20,moehle21}
we proposed novel enumeration and counting 
 procedures based on entailment. 
Unfortunately we were not able to produce any 
implementation efficient enough to compete with current enumeration procedures based
on \verbverification.
Only very recently, in \cite{friedEntailingGeneralizationBoosts2024} we
 have enhanced the {\sf HALL} enumeration tool for circuits from
 \cite{friedAllSATCombinationalCircuits2023} by substituting
 \verbverification{} checks with entailment checks for
 partial-assignment reductions (``generalizations''
 in \cite{friedEntailingGeneralizationBoosts2024}),
boosting its performance in terms of both
time efficiency and size of partial-assignment sets.

In \cite{masinaCNFConversionDisjoint2023} we addressed a problem which
is different although related to $(e)$, showing that adopting a form 
of CNF-ization different from
\cite{tseitin1,plaistedStructurepreservingClauseForm1986} improves the
efficiency and effectiveness of enumeration. Such
encoding, however, does not fix fact $(e)$ (see \sref{sec:partialsat-cnf}).

\paragraph{Content.} The rest of the paper is organized as follows.
\sref{sec:background} provides the necessary notation, terminology, and concepts used in the paper.
\sref{sec:partialsat} presents our theoretical analysis:
\sref{sec:partialsat-plain}
introduces \verbverification{} and entailment
 for generic propositional formulas and discusses their
 relative properties;
\sref{sec:partialsat-exist} lifts the discussion to
existentially-quantified formulas;
\sref{sec:partialsat-cnf} discusses the role of CNF-ization in this
context;
 \sref{sec:partialsat-others} analyzes other candidate forms of partial-assignment satisfaction.
\sref{sec:practical} discusses the practical consequences
of entailment vs. \verbverification{}.
\sref{sec:concl} provides 
conclusions and suggestions for future research.

\section{Background}
\label{sec:background}

In this section we introduce the 
notation and terminology adopted in this paper, and 
we recall the standard
syntax, semantics and basic facts of propositional logics.

%

\paragraph{Notation.}
In what follows
$\tval$, $\fval$, $\uval$ denote the
truth values ``true'', ``false'' and ``unknown'' respectively;
$\top$, $\bot$ denote the logic constants
``true'' and ``false'' respectively; 
%
$A$, $B$ denote propositional atoms;
%
$\varphi$, $\psi$ denote propositional formulas;
%
$\mu,\eta,\gamma$ denote truth value assignments.
%
The symbols $\alla\defas\set{A_1,...,A_N}$ and
$\allb\defas\set{B_1,...,B_K}$ 
denote disjoint sets of propositional atoms. 
More precisely, $\vi$ and $\psi$ denote generic propositional formulas
built on $\alla$, $\allb$ and $\alla\cup\allb$ respectively;
 $\eta$ and $\mu$ denote total and a partial
assignments on $\alla$ respectively;
$\delta$ denote total assignments on $\allb$. (All above symbols may possibly
have subscripts.)

\paragraph{Syntax.}
A \new{propositional formula} is defined inductively as follows: 
the constants $\top$ and $\bot$ (denoting the truth values true and false)
are formulas; 
   a {propositional atom} 
   $A_1,A_2,A_3,...$ is a formula;
   if $\varphi_1$ and $\varphi_2$ are formulas, then 
\ignore{ $\neg\varphi_1$ and 
  $\varphi_1\bowtie\varphi_2$ are formulas, s.t. $\bowtie\ \in\set{\wedge,\vee,\imp,\limp,\iff}$.
}
  $\neg\varphi_1$ and 
  $\varphi_1\wedge\varphi_2$
%
  are formulas.
We use the standard Boolean abbreviations:  
``$\varphi_1\vee\varphi_2$'' for  
``$\neg(\neg\varphi_1\wedge\neg\varphi_2)$'', 
``$\varphi_1\imp\varphi_2$'' for ``$\neg(\varphi_1\wedge\neg\varphi_2)$'', 
``$\varphi_1\iff\varphi_2$'' for 
``$\neg(\varphi_1\wedge\neg\varphi_2) \wedge
\neg(\varphi_2\wedge\neg\varphi_1)$''.
%
A \new{literal} is either an atom (a \new{positive
 literal}) or its negation (a \new{negative
 literal}).
(If $l$ is a negative literal $\neg A_i$,
then by ``$\neg l$'' we conventionally mean $A_i$ rather than
$\neg\neg A_i$.)
A \new{clause} is a disjunction of literals $\bigvee_j l_j$.
A \new{cube} is a conjunction of literals $\bigwedge_j l_j$.
$\varphi$ is in \new{Conjunctive Normal Form (CNF)} iff it is
   a conjunction of clauses:
$
\bigwedge_{i=1}^{L}\bigvee_{j_i=1}^{K_i} l_{j_i}
$.

\paragraph{Semantics.}
Given $\alla\defas\set{A_1,...,A_N}$,
a map  $\eta:\alla\longmapsto\set{\trueval,\falseval}$ is a
\new{total truth assignment} for $\alla$. 
{We assume $\eta(\top)\defas \trueval$ and $\eta(\bot)\defas \falseval$.}
We  represent $\eta$ as a \new{set of literals}
$\eta\defas \set{A_i\ |\ \eta(A_i)=\trueval}\cup\set{\neg A_i\ |\
  \eta(A_i)=\falseval}$. We sometimes represent $\eta$ also as a
\new{cube} 
$
\bigwedge_{\eta(A_i)=\tiny{\trueval}} A_i \wedge 
\bigwedge_{\eta(A_i)=\tiny{\falseval}}\neg A_i
$ which we denote as ``$\andeta$'' so that to distinguish the set and the cube
representations. We denote by $|\eta|$ the number of literals in $\eta$.

Given a {\em total} truth assignment $\eta$ on \alla{} and some formulas
$\vi,\vi_1,\vi_2$ on \alla, the sentence ``$\eta$ {\em satisfies} $\varphi$'', written
``$\eta\models\varphi$'', 
is defined recursively on the structure of $\varphi$ as follows:
$\eta\models \top$, 
$\eta\not\models \bot$, 
$\eta\models A_i$ if and only if $\eta(A_i)=\trueval$,
$\eta\models\neg\vione$ if and only if $\eta\not\models\vione$,
$\eta\models\vione\wedge\vitwo$ if and only if
$\eta\models\vione$ and $\eta\models\vitwo$.
{(The definition of $\eta\models\vi_1\bowtie\vi_2$ for the other
  connectives  follows straightforwardly from their definition in terms
  of $\neg,\wedge$.)}
%
$\varphi$ is \new{satisfiable} iff $\eta\models\varphi$ for some total truth
assignment $\eta$ on $\alla$.
$\varphi$ is \new{valid} (written ``$\models \vi$'') iff $\eta\models\varphi$ for every total truth
assignment $\eta$ on $\alla$. 
\new{$\varphi_1$ entails $\varphi_2$} (written \new{``$\varphi_1\entails
     \varphi_2$''}) iff,  for every total assignment $\eta$ on
   $\alla$, if
 $\eta\models\varphi_1$
then $\eta\models\varphi_2$.
$\varphi_1$ and $\varphi_2$ are \new{ equivalent} (written \new{``$\varphi_1\equiv \varphi_2$''}) iff 
$\varphi_1\entails  \varphi_2$ and $\varphi_2\entails  \varphi_1$.
Consequently: $\varphi$ is unsatisfiable iff $\neg\varphi$ is valid;
$\vione\models\vitwo$ iff $\vione\imp\vitwo$ is valid;
a clause $\bigvee_i l_i$ is valid (aka is a \new{tautology}) iff both $A_i$ and $\neg A_i$
occur in it for some $A_i$; a CNF formula \vi is valid iff either it
is $\top$ or all its clauses are tautologies. 
We say that a CNF formula is \new{tautology-free} iff none of its clauses
is a tautology.
%

%
\begin{figure}[t!]
  \small
  \begin{tabular}{ccc}
  \begin{minipage}[t]{0.5\textwidth}
    $
    \begin{array}{|cl|cl|}
  \hline    
      \neg\top & \Rightarrow \bot & \neg \bot & \Rightarrow \top \\
      \top\wedge\vi, \vi\wedge\top  & \Rightarrow \vi & \bot\wedge\vi,  \vi\wedge\bot& \Rightarrow \bot \\
      \top\vee\vi, \vi\vee\top& \Rightarrow \top & \bot\vee\vi, \vi\vee\bot& \Rightarrow \vi \\
      \top\imp\vi & \Rightarrow \vi & \bot\imp\vi & \Rightarrow \top \\
      \vi\imp\top & \Rightarrow \top & \vi\imp\bot & \Rightarrow \neg\vi \\
      \top\iff\vi, \vi\iff\top& \Rightarrow \vi & \bot\iff\vi, \vi\iff\bot& \Rightarrow \neg\vi \\
  \hline         
    \end{array}
    $\\ 
    \caption{\label{fig:boolprop}
    Propagation of truth values 
through the Boolean connectives.  }
\end{minipage}%
      & \hspace{.025\textwidth} &
\begin{minipage}[t]{0.45\textwidth}
    $
      \begin{array}{||c||l|l|l|l|l|l|l|l|l||}
  \hline    
  \muof{\vi_1}            & \tval & \tval & \tval & \uval & \uval & \uval &\fval & \fval & \fval\\
  \muof{\vi_2}            & \tval & \uval & \fval & \tval & \uval & \fval &\tval & \uval & \fval\\
     \hline
  \muof{\neg\vi_1}        & \fval & \fval & \fval & \uval & \uval & \uval &\tval & \tval & \tval\\
  \muof{\vi_1\wedge\vi_2} & \tval & \uval & \fval & \uval & \uval & \fval &\fval & \fval & \fval\\
  \muof{\vi_1\vee\vi_2}   & \tval & \tval & \tval & \tval & \uval & \uval &\tval & \uval & \fval\\
  \muof{\vi_1\imp\vi_2}   & \tval & \uval & \fval & \tval & \uval & \uval &\tval & \tval & \tval\\
  \muof{\vi_1\iff\vi_2}   & \tval & \uval & \fval & \uval & \uval & \uval &\fval & \uval & \tval\\
     \hline
      \end{array}
      $
\caption{\label{fig:threeval}
Three-value-semantics of $\muof{\vi}$ in terms of \set{\tval,\fval,\uval}%
.\newline
\ignoreinlong{The definition of
$\muof{\vi_1\bowtie\vi_2}$ s.t. $\bowtie\ \in\set{\vee,\imp,\limp,\iff}$
follows straighforwardly.}
}
\end{minipage}
\end{tabular}
\end{figure}

\paragraph{Partial truth assignments.}
A map  $\mu:\alla'\longmapsto\set{\trueval,\falseval}$
s.t. $\alla'\subseteq\alla$ and $N'\defas|\alla'|$ is a
\new{partial truth assignment} for $\alla$. 
%
%
As with total assignments, we can represent $\mu$ as a set of literals or as
a cube, denoted with ``$\andmu$''. 

By ``{\em apply a partial assignment $\mu$ to $\varphi$}'' we mean 
``substitute all instances of each assigned
$A_i$ in $\varphi$ with the truth constants in $\set{\top,\bot}$
corresponding to the truth value assigned by
$\mu$, and then apply recursively the standard
propagation of truth constants through the Boolean connectives
described in Figure~\ref{fig:boolprop}.
%
We denote by ``$\applymuvi$'' (``{\em residual of $\vi$ under $\mu$''}) the formula resulting from applying
$\mu$ to $\vi$.

We introduce a three-value logic to extend $\mu$ to \alla{} as
$\mu:\alla\longmapsto\set{\trueval,\falseval,\unknown}$
by assigning to \unknown{} (unknown) the unassigned atoms in $\alla\setminus \alla'$.
Then we extend the semantics of $\mu$ to any formula $\vi$ on \alla as
described in Figure~\ref{fig:threeval}.

The following facts follow straightforwardly and are of interest for our discussion.

\begin{Property}
\label{prop:total-properties}
Let $\eta$ be a total truth assignment on \alla{} and
$\vi,\vione,\vitwo$ be formulas on \alla. 
\begin{renumerate}
\item 
$\eta\models\varphi$ iff $\andeta\entails\varphi$.
\item 
If $\vione$ and $\vitwo$ are  equivalent, then 
$\eta\models\vione$ iff $\eta\models\vitwo$.
\item
$\eta\models\varphi$ iff \applyetavi{} is $\top$. 
\item 
Checking
 $\eta\models\vi$
requires at most a polynomial amount of steps.  
\end{renumerate}  
\end{Property}
\noindent
Notice that Property~\ref{prop:total-properties}(i) justifies the
  usage of ``$\models$'' 
for both satisfiability and entailment.

\begin{Property}
  \label{prop:eval-simplifies}
Let $\mu$ be a partial assignment on \alla{} and
$\vi$ be a formula on \alla. \\Then
$\applymuvi$ is $\top$ iff $\muof{\vi}=\tval$ and
$\applymuvi$ is $\bot$ iff $\muof{\vi}=\fval$.
\end{Property}
\begin{IGNORE}%
  \begin{Property}%
  $\andmu\wedge\applymuvi\ \equiv\ \andmu\wedge\vi$.
\end{Property}  
\end{IGNORE}

\noindent
Notice that total assignments are a subcase of partial ones, so that
the definition of residual and Property~\ref{prop:eval-simplifies} 
apply also to total assignments $\eta$.

\paragraph{Existentially-quantified formulas.}
%
A total truth assignment $\eta$ satisfies $\exists \allb. \psi$,
written ``$\eta\models\exists \allb. \psi$'',
iff exists a total truth assignment $\delta$ on \allb s.t.
$\eta\cup\delta\models\psi$.
We call the {\em Shannon expansion} \shannonpsib of the existentially-quantified
formula $\exists \allb. \psi$ the propositional formula on \alla
defined as 
\begin{eqnarray}
\label{eq:shannon}
\shannonpsib\defas 
\textstyle
\bigvee_{\delta_i\in \set{{\scriptsize \tval,\fval}}^{|\allb|}}
\apply{\psi}{\delta_i}
\end{eqnarray}
where $\set{{\tval,\fval}}^{|\allb|}$ is the set of all total
assignments on \allb.
{Notice that some $\apply{\psi}{\delta_i}$ may be inconsistent or $\bot$.}
The following property derives directly from the above definitions.

\begin{Property}
\label{prop:total-shannon}
Let   $\psi$ be a formula on
$\alla\cup\allb$ and $\eta$ be a total truth assignment on \alla.
\\Then
$\eta\models\exists \allb. \psi$\ iff\ $\eta\models\shannonpsib$,
that is, $\exists \allb. \psi\equiv \shannonpsib $.
\end{Property}

\paragraph{CNF-ization.}
Every generic formula $\vi$ on \alla can be encoded into a CNF formula $\psi$
on $\alla\cup\allb$ for some \allb by applying (variants of) Tseitin
CNF-ization \TseitinCNF(\vi) \cite{tseitin1}, consisting in applying recursively bottom-up 
the rewriting rule:
\begin{eqnarray}
\label{eq:tseitin}
\vi &\Rightarrow&
\vi[(l_{j1} \bowtie l_{j2})\mapsto B_j]\wedge \DeMorganCNF(B_j \iff (l_{j1} \bowtie l_{j2}))  
\end{eqnarray}
until the resulting formula $\psi$ is in CNF, where $l_{j1},l_{j2}$ are
literals, $\bowtie\ \in\ \set{\wedge,\vee,\imp,\limp,\iff}$ and 
$\DeMorganCNF()$ is the validity-preserving CNF conversion based on DeMorgan
rules (e.g., 
$\DeMorganCNF(B \iff (l_1 \wedge l_2)) \defas 
(\neg B \vee l_1) \wedge 
(\neg B \vee l_2) \wedge 
(B \vee \neg l_1 \vee \neg l_2)$).
The size of $\psi$ is linear wrt. that of $\vi$.  
With Plaisted\&Greenbaum CNF-ization \PlaistedCNF(\vi)
\cite{plaistedStructurepreservingClauseForm1986}
the term $B_j \iff (l_{j1} \bowtie l_{j2})$ in \eqref{eq:tseitin} is
substituted with $B_j \imp (l_{j1} \bowtie l_{j2})$ \resp{$B_j \limp
  (l_{j1} \bowtie l_{j2})$} if $(l_{j1} \bowtie l_{j2})$ occurs only
positively \resp{negatively} in $\vi$.

\begin{Property}{}
\label{prop:cnfequiv}
If $\psi\defas\TseitinCNF(\vi)$ or $\psi\defas\PlaistedCNF(\vi)$, then
$\eta\models\vi$ iff exists a total assignment $\delta$ on
\allb s.t. $\eta\cup\delta\models\psi$, that is,
$\vi\equiv\exists\allb.\psi$.
\end{Property}


\section{A theoretical analysis of \verbverification{} and entailment}
\label{sec:partialsat}

We address the following basic question:
{\em What should we mean by ``a {partial} assignment $\mu$ satisfies $\vi$"?}
We wish to provide a satisfactory definition of partial-assignment
satisfiability for a generic propositional formula ---i.e., not
only for (tautology-free) CNF. 
Ideally, a suitable definition of partial-assignment satisfiability
should verify all statements in Property~\ref{prop:total-properties},
in particular (ii) and (iv). 
In practice, unfortunately, at least for generic (i.e., non-CNF) formulas, we show
 this is not the case. 

\subsection{\Verbverification{} and entailment of formulas.}
\label{sec:partialsat-plain}

\begin{IGNORE}%
We use a three-value logic to extend $\mu$ to \alla{} as
$\mu:\alla\longmapsto\set{\trueval,\falseval,\unknown}$
by assigning to \unknown{} (unknown) the unassigned atoms in $\alla\setminus \alla'$.
Then we extend the semantics of $\mu$ to any formula $\vi$ on \alla as
described in Figure~\ref{fig:threeval}. 
\begin{Property}
  \label{prop:eval-simplifies}
Let $\mu$ be a partial assignment on \alla{} and
$\vi$ be a formula on \alla. \\Then
$\applymuvi$ is $\top$ iff $\muof{\vi}=\tval$ and
$\applymuvi$ is $\bot$ iff $\muof{\vi}=\fval$.
\end{Property}
\end{IGNORE}

\paragraph{Definitions.}
The first candidate definition of partial-assignment satisfaction,
which extends to partial assignments 
Property~\ref{prop:total-properties}(iii),
is 
{\em \verbverification}. 
\begin{definition}
\label{def:verification}
We say that a {\em partial} truth assignment $\mu$  {\bf \verbverifies{}} $\varphi$ 
iff $\muof{\vi}=\tval$ ---or, equivalently by
Property~\ref{prop:eval-simplifies}, iff $\applymuvi=\top$.
We denote this fact with ``$\mu\verifies\varphi$''.
\end{definition}

\noindent
We stress the fact that \verbverification{} is a {\em semantic}
definition ($\muof{\vi}=\tval$) which, due to 
Property~\ref{prop:eval-simplifies},
captures exactly the {\em syntactic} and easy-to-check satisfaction criterion ($\applymuvi=\top$) which is implicitly used in many procedures. 
%

The second candidate definition of partial-assignment satisfaction, which extends to partial assignments 
Property~\ref{prop:total-properties}(i),
is {\em entailment}. 

\begin{definition}
\label{def:entailment}
  We say that a {\em partial} truth assignment $\mu$  {\bf entails} $\varphi$
if and only if, for every total truth assignments $\eta$
s.t.$\mu\subseteq\eta$, 
$\eta$ satisfies $\varphi$ ---or, equivalently, iff $\applymuvi$ is
valid. 
We denote this fact with ``$\mu\entails\varphi$''.
\end{definition}

%

\paragraph{\Verbverification{} vs. Entailment.}
We show the following facts. 
When the formula $\vi$ is a tautology-free CNF
then
\verbverification{} and entailment
coincide: 
$\mu\verifies\varphi$ iff $\mu\entails\varphi$.
\ignore{
  In fact, if $\mu\verifies\vi$ then,  for every $\eta$ s.t.
$\eta\supseteq\mu$, $\eta\verifies\vi$ and thus $\eta\models\vi$, hence
$\mu\entails\vi$;
also, if $\mu\entails\vi$ then $\applymuvi$ is a valid
  CNF formula which does not contain valid clauses, so that $\applymuvi$
  must be $\top$, hence $\mu\verifies\vi$.
}
\ignore{%
(We conjecture that this is the reason why the distinction between
these two concepts 
has been long overlooked in the literature.)}
%
Unfortunately, \emph{with generic 
formulas 
\verbverification{} and entailment
  do not coincide, the former being
strictly stronger than the latter}. 
%

\begin{theorem}
\label{prop:stronger}
\label{prop:equivalentifcnf}
Let \vi be a formula and $\mu$ be a partial truth
assignment over its atoms.
\begin{itemize}
\item[(a)] If \vi{} a tautology-free CNF, then
$\mu\entails\vi$ iff $\mu\verifies\vi$.
\item[(b)] If \vi{} is a generic formula, then
$\mu\entails\vi$ if $\mu\verifies\vi$, but the converse does not hold in general.
\end{itemize}

\end{theorem}
\begin{proof}
$(a)$: Let $\vi$ be a tautology-free CNF. 
If $\mu\entails\vi$, then $\applymuvi$ is a valid
  CNF formula which does not contain valid clauses, so that
  $\applymuvi=\top$, hence $\mu\verifies\vi$  by
  Property~\ref{prop:eval-simplifies}.\\
$(a)$ and $(b)$:
If $\mu\verifies\vi$ then,  for every $\eta$ s.t.
$\eta\supseteq\mu$, $\eta\verifies\vi$ and thus $\eta\models\vi$, hence
$\mu\entails\vi$.\\
$(b)$: The fact that the converse does not hold in general is shown in \cref{ex:stronger}.
\hfill $\qed$
\end{proof}
  \begin{example}
  \label{ex:stronger}
If $\vi\defas(A_1\wedge A_2)\vee (A_1\wedge \neg A_2)$ and 
$\mu\defas\set{A_1}$,
then 
$\mu\entails\vi$ but $\mu\not\verifies\vi$. $\diamond$
\end{example}
\noindent
Hereafter we implicitly assume w.l.o.g. that all CNF formulas are tautology free.

We try to build a counterpart of Property~\ref{prop:total-properties}
for Definitions~\ref{def:verification} and
\ref{def:entailment} respectively, but in both cases we fail to
achieve all points (i)-(iv), 
resulting into complementary situations.
%
The following properties follow straightforward from Definitions~\ref{def:verification} and  \ref{def:entailment}.
(Here ``\oksym{}'' \resp{``\noksym{}''} denotes facts from
Property~\ref{prop:total-properties} which are \resp{are not} preserved.)

\begin{Property}
\label{prop:partial-verification}
Let $\mu$ be a partial truth assignment on \alla{} and
$\vi,\vione,\vitwo$ be formulas on \alla. 
\begin{renumerate}
\item \noksym{}
If $\mu\verifies\varphi$ then $\andmu\entails\varphi$, {but not vice versa}.
\item \noksym{}
If $\vione$ and $\vitwo$ are  equivalent, this {does not} imply
that $\mu\verifies\vione$ iff $\mu\verifies\vitwo$.
\item \oksym{}
  $\mu\verifies\varphi$ iff \applymuvi{} is $\top$
     (also, iff $\muof{\vi}= \tval$ by Property~\ref{prop:eval-simplifies}). 

\item \oksym{}
Checking
$\mu\verifies\vi$  requires at most a polynomial amount of steps.
\end{renumerate}  
\end{Property}

\ignore{From Definition~\ref{def:entailment}
we easily derive the following.}

\begin{Property}
\label{prop:partial-entailment}
Let $\mu$ be a partial truth assignment on \alla{} and
$\vi,\vione,\vitwo$ be formulas on \alla. 
\begin{renumerate}
\item \oksym{}
$\mu\entails\varphi$ iff $\andmu\entails\varphi$.
\item \oksym{}
If $\vione$ and $\vitwo$ are  equivalent, then 
$\mu\entails\vione$ iff $\mu\entails\vitwo$.
\item \noksym{}
  $\mu\entails\varphi$ iff \applymuvi{} is a valid formula, not
  necessarily $\top$
   (also, in general $\muof{\vi} \neq \tval$). 
\item \noksym{}
Checking
$\mu\entails\vi$ is \emph{co-NP-complete}.
\ignore{
  \footnote{In fact, checking the validity of \vi{}
translates into verifying that the empty assignment
entails it.}}
\end{renumerate}  
\end{Property}

\begin{example}
  \label{ex:embarassing}
  Let $\vi_1\defas(A_1\wedge A_2)\vee (A_1\wedge \neg A_2)$,
  $\vi_2\defas A_1$ and 
$\mu\defas\set{A_1}$.
Then
$\vi_1\equiv\vi_2$, $\apply{\vi_1}{\mu} = A_2\vee\neg A_2\neq \top$ and $\apply{\vi_2}{\mu}=\top$.
Thus
$\andmu\entails\vi_1$,
$\mu\entails\vi_1$ and $\mu\entails\vi_2$,
whereas
$\mu\not\verifies\vi_1$ but $\mu\verifies\vi_2$.
\hfill $\diamond$
\end{example}

\ignore{From a theoretical viewpoint, \cref{ex:embarassing} spotlights the difference between
Property~\ref{prop:partial-verification}(ii) and
Property~\ref{prop:partial-entailment}(ii)
(see also  Property~\ref{prop:total-properties}(ii)):}
From a theoretical viewpoint, \cref{ex:embarassing} spotlights the
difference between
Property~\ref{prop:partial-verification}(i) and
Property~\ref{prop:partial-entailment}(i) and, even more importantly, that between
Property~\ref{prop:partial-verification}(ii) and
Property~\ref{prop:partial-entailment}(ii):
\emph{whereas entailment  matches the intuition 
that equivalent 
formulas should be satisfied by the same (partial) assignments,
\verbverification{} does not}. 
The latter fact looks theoretically awkward. 
%
%
In particular, 
if we adopted \verbverification{} as the definition of  partial-assignment
satisfiability,
then  we believe that it would be embarrassing to have that equivalent formulas were not ``satisfied''
by the same assignments.



\subsection{\Verbverification{} and entailment of
  existentially-quantified formulas.}
\label{sec:partialsat-exist}

We extend the analysis of \sref{sec:partialsat-plain} to
existentially-quantified propositional formulas, 
 wishing to provide a satisfactory definition of partial-assignment
 satisfiability for this case as well.
Property~\ref{prop:total-shannon} paves our way, telling how to
extend the 
definitions of \verbverification{} and entailment
to existentially-quantified formulas. 


The first possibility is to see partial-assignment satisfiability as
{\em \verbverification}, leveraging Definition~\ref{def:verification}
to the existentially-quantified
case by exploiting  Shannon expansion.
%
%
\begin{definition}
\label{def:verification-exist}
 We say that a {\em partial} truth assignment $\mu$ on $\alla$ {\bf
   \verbverifies} $\exists
\allb.\psi$ if and only if, there exists a total truth assignment 
$\delta$ on $\allb$ s.t.
\ignore{the result of applying $\mu\cup\delta$ to 
$\psi$ is $\top$, that is,}
$\mu\cup\delta\verifies\psi$, that is, $\apply{\psi}{\mu\cup\delta}=\top$.
\end{definition}

\begin{theorem}
\label{prop:verification-exist}
Let   $\psi$ be a formula on
$\alla\cup\allb$ and $\mu$ be a partial assignment on \alla.
Then\\
  $\mu\verifies\exists \allb. \psi$ iff $\mu\verifies\shannonpsib$. 
\end{theorem}

\begin{proof}
 By Definition~\ref{def:verification},  $\mu\verifies{} \shannonpsib$ iff
 $\apply{(\shannonpsib)}{\mu} = \top$, 
 that is, 
 by \eqref{eq:shannon},
 iff there exists some
 $\delta_i$ s.t. $\apply{\apply{\psi}{\delta_i}}{\mu}=\top$ (i.e., $\apply{\psi}{\delta_i\cup\mu}=\top$),
 that is, by Definition~\ref{def:verification}, iff there exists some $\delta_i$ s.t. $\mu\cup\delta_i\verifies\psi$,
 that is, by Definition~\ref{def:verification-exist}, iff
 $\mu\verifies\exists \allb. \psi$.
 \hfill $\qed$
\end{proof}
\noindent Notice  that \cref{prop:verification-exist} is
{\em not} a straighforward consequence of
Property~\ref{prop:total-shannon}, because equivalent formulas are not
necessarily verified by the same assignments
(Property~\ref{prop:partial-verification}(ii)). 

One second possibility is to see partial-assignment satisfiability as
{\em entailment}, leveraging Definition~\ref{def:entailment}
and Property~\ref{prop:total-shannon} to the existentially-quantified
case. %
This leads to the following definition and relative Property. 
\begin{definition}
\label{def:entailment-exist}
We say that a {\em partial} truth assignment $\mu$ on $\alla$ {\bf entails}
$\exists \allb.\psi$, written $\mu\entails\exists \allb.\psi$,
if and only if , for every total truth assignment $\eta$ on $\alla$ 
extending $\mu$,
there exists a total truth assignment $\delta$ on $\allb$
s.t. $\eta\cup\delta$ satisfies $\psi$. 
\end{definition}

\begin{theorem}
\label{prop:entailment-exist}
Let   $\psi$ be a formula on
$\alla\cup\allb$ and $\mu$ be a partial assignment on \alla.
Then\\
  $\mu\entails\exists \allb. \psi$ iff $\mu\entails\shannonpsib$. 
\end{theorem}

\begin{proof}
  $\mu\entails \shannonpsib$ iff $\eta\models\shannonpsib$ for every total
assignment $\eta$ s.t. $\eta\supseteq\mu$, 
that is, by Property~\ref{prop:total-shannon}, iff for every total
assignments $\eta$ s.t. $\eta\supseteq\mu$ exists a total assignment 
$\delta$ on \allb s.t. $\eta\cup\delta\models\psi$, that is, iff
$\mu\entails\exists \allb. \psi$. \hfill $\qed$
\end{proof}

  \noindent
  Notice the nesting order of forall/exists in
Definition~\ref{def:entailment-exist}: ``for every $\eta$ exists $\delta$
s.t. ...''. In fact, distinct $\eta$'s may satisfy distinct disjuncts 
$\apply{\psi}{\delta_i}$ in \shannonpsib{}, requiring distinct 
$\delta_i$'s.
\ignore{Therefore, we notice that the following statement, inverting the
quantification order, is \emph{not} a suitable definition for $\mu\entails\exists \allb.\psi$:
%
\\ \noindent
  ``{\em We say that a {\em partial} truth assignment $\mu$ on $\alla$ entails
  $\exists \allb.\psi$ if and only if, there exists a total truth
  assignment $\delta$ on $\allb$ s.t., for every total truth assignment
  $\eta$ extending $\mu$, $\eta\cup\delta$ satisfies $\psi$.}''
}
%

\begin{theorem}
\label{prop:stronger-exists}
Let $\psi$ be a formula on $\alla\cup\allb$ and $\mu$ be a partial truth
assignment over $\alla$.
Then
$\mu\entails\exists\allb.\psi$ if $\mu\verifies\exists\allb.\psi$, but
the converse does not hold in general. 
\end{theorem}
\begin{proof}
  From
  \cref{prop:stronger}$(b)$ with $\vi\defas \shannonpsib{}$
  we can deduce that $\mu\entails\shannonpsib{}$ if
  $\mu\verifies\shannonpsib{}$.
By \cref{prop:verification-exist} and
\cref{prop:entailment-exist} we have that $\mu\entails\exists\allb.\psi$ if $\mu\verifies\exists\allb.\psi$.
\\The fact that the converse does not hold in general is shown in \cref{ex:exists-stronger}.
\hfill $\qed$
\end{proof}

\noindent

\begin{example}
\label{ex:exists-stronger}
  Consider $\mu\defas\set{A_1}$
  and the tautology-free CNF formula on $\alla\cup\allb$:\\
$
\begin{array}{lll}
\psi\defas  
& (B_1\vee B_2)\ \wedge & \\
& (\neg B_1 \vee \pos A_1) \wedge
(\neg B_1 \vee \pos A_2) \wedge
(B_1\vee \neg A_1 \vee \neg A_2)\ \wedge\ & \ \ \ \\
& (\neg B_2 \vee \pos A_1) \wedge
(\neg B_2 \vee \neg A_2) \wedge
(B_2\vee \neg A_1 \vee \pos A_2).\ & \ \ \ 
\\
  \shannonpsib= &
  \xcancel{(A_1\wedge A_2\wedge\neg A_2)}\vee (A_1\wedge A_2)\vee (A_1\wedge\neg A_2) 
\vee \xcancel{\bot}.\\
\end{array}
$\\
Thus $\mu\entails\shannonpsib$ but 
$\mu\not\verifies\shannonpsib$,
so  that $\mu\entails\exists\allb.\psi$ but
$\mu\not\verifies\exists\allb.\psi$.
\hfill $\diamond$
\end{example}

Thus
\verbverification{} is strictly stronger than entailment also with
existentially-quantified formulas.
Remarkably, and unlike with the un-quantified case, this is the case 
 \emph{even if $\psi$ is a tautology-free CNF formula}! (See \cref{ex:exists-stronger}.)
Intuitively, this can be seen as a consequence of the fact that 
\shannonpsib is not in CNF even if $\psi$ is in CNF. 




\subsection{\Verbverification{} and entailment of CNF-ized non-CNF formulas.}
\label{sec:partialsat-cnf}

One might argue that in SAT-related problems the distinction between verification and
entailment in \sref{sec:partialsat-plain} is not relevant in practice, because we could
CNF-ize upfront the input non-CNF formulas and then remove tautological clauses,
exploiting Property~\ref{prop:cnfequiv} and the fact that the above
distinction does not hold for (tautology-free) CNF
formulas.

Unfortunately, the following result shows this is not the case.
%

\begin{theorem}
\label{prop:cnf-stronger}
Let $\vi$ be a non-CNF formula on $\alla$ and $\psi\defas
\TseitinCNF(\vi)$ or $\psi\defas
\PlaistedCNF(\vi)$,
and let $\mu$ be a partial truth
assignment over $\alla$.
Then
$\mu\entails\exists\allb.\psi$ if $\mu\verifies\exists\allb.\psi$, but
the converse does not hold in general. 
\end{theorem}

\begin{proof}
As an instantiation of \cref{prop:stronger-exists} we have 
$\mu\entails\exists\allb.\psi$ if $\mu\verifies\exists\allb.\psi$.
  \\The fact that the converse does not hold in general is shown in
  \cref{ex:exists-cnf-stronger}.
  \hfill $\qed$
\end{proof}

\begin{example}
\label{ex:exists-cnf-stronger}
Consider $\vi\defas (A_1\wedge A_2) \vee (A_1\wedge \neg A_2)$ and
$\mu\defas\set{A_1}$.
Then $\psi\defas\TseitinCNF(\vi)$ 
is the formula in \cref{ex:exists-stronger}.
Then $\mu\entails\exists\allb.\TseitinCNF(\vi)$ but
$\mu\not\verifies\exists\allb.\TseitinCNF(\vi)$.\\
Let $\psi\defas\PlaistedCNF(\vi)$ instead. Since $\psi$ is \TseitinCNF(\vi) minus
its two ternary clauses, we also have that
$  \shannonpsib= 
  \xcancel{(A_1\wedge A_2\wedge\neg A_2)}\vee (A_1\wedge A_2)\vee (A_1\wedge\neg A_2) 
\vee \xcancel{\bot}.$\\
Thus $\mu\entails\exists\allb.\PlaistedCNF(\vi)$ but
$\mu\not\verifies\exists\allb.\PlaistedCNF(\vi)$.
\hfill $\diamond$
\end{example}

Intuitively, the whole point is that both \TseitinCNF(\vi) and \PlaistedCNF(\vi)
introduce fresh variables \allb{}, which must be implicitly existentially
quantified away in order to preserve equivalence and thus produce the set
of original satisfying assignments (Property~\ref{prop:cnfequiv}).
Although these variables do not affect entailment, they may affect
\verbverification{}. In fact, since $\vi$ and $\exists\allb.\TseitinCNF(\vi)$
\resp{$\exists\allb.\PlaistedCNF(\vi)$} are equivalent, they are
entailed by the same partial assignments, but they are not
\verbverified{} by the same partial assignments.

To this extent, in \cite{masinaCNFConversionDisjoint2023} we addressed
a different though related problem: we proved 
that ---unlike with $\psi\defas\TseitinCNF(\vi)$ or with
$\psi\defas\PlaistedCNF(\vi)$--- if $\psi\defas\PlaistedCNF(\NNF{\vi})$ and 
$\mu\verifies\vi$, then $\mu\verifies\exists\allb.\psi$.
Notice, however, that this fact does not apply to our problem, 
because \cref{prop:cnf-stronger} holds also for $\psi\defas\PlaistedCNF(\NNF{\vi})$.

\begin{example}
  Consider $\vi$ and $\mu$ as in \cref{ex:exists-cnf-stronger}.
  Since $\vi$ is already in NNF, then we have that
  $\psi\defas\PlaistedCNF(\NNF{\vi}) = \PlaistedCNF(\vi)$.
  Therefore $\mu\entails\exists\allb.\PlaistedCNF(\NNF{\vi})$ but
$\mu\not\verifies\exists\allb.\PlaistedCNF(\NNF{\vi})$.
\hfill $\diamond$

\end{example}


\subsection{Other candidate forms of partial-assignment satisfaction.}
\label{sec:partialsat-others}
Given the above facts, we wonder if we could use other notions
of partial-assignment satisfiability. (E.g., we could enrich
$\verifies$ with some form of formula simplification.)
In particular, we wonder if there exists some suitable notion of
partial-assignment satisfaction which is strictly stronger then 
entailment and fixes the ``embarrassing fact'' for \verbverification{} of
Property~\ref{prop:partial-verification}(ii).
The following theorem states this is not the case.

\begin{theorem}
  \label{uniqueness:teo}
  Let $\genericentails$ denote some relation
  such that
\begin{aenumerate}
\item for every $\mu$ and $\vi$,  $\mu\entails\vi$ if
$\mu\genericentails\vi$;
\item{\label{item:total}}
   for every $\eta$ and $\vi$,
   if $\eta$ is total for the set of atoms in $\vi$, then
   $\eta\genericentails\vi$ iff
   $\eta\models\vi$;
 \item{\label{item:and}} for every $\mu$, $\vi_1$ and $\vi_2$,
   if $\mu\genericentails\vi_1\wedge\vi_2$, then $\mu\genericentails\vi_1$ and
  $\mu\genericentails\vi_2$;
\item for every $\mu$, $\vi_1$ and $\vi_2$,
  if $\vi_1\equiv\vi_2$, then $\mu\genericentails\vi_1$ iff
  $\mu\genericentails\vi_2$
\end{aenumerate}
Then, for every $\mu$, $\vi$, we have that $\mu\genericentails\vi$ if and only if $\mu\entails\vi$.
\end{theorem}
\begin{proof}
  The ``only if'' part  is hypothesis {\em (a)}.\\
  The ``if'' part: By absurd, suppose there exist $\mu$ and $\vi$ s.t. $\mu\entails\vi$ and $\mu\not\genericentails\vi$ .
  \\Let 
  $\vi_1\defas\andmu$ and $\vi_2\defas\andmu\wedge\varphi$.
  Then we have that:
  \\$\vi_1\equiv\vi_2$, because $\mu\entails\vi$, and hence $\andmu\entails\vi$ because of
  Property~\ref{prop:partial-entailment}(i);
  \\$\mu\genericentails\vi_1$,
  because of hypothesis {\em (b)};
  \\$\mu\not\genericentails\vi_2$, because of hypothesis {\em
    (c)}.
  \\ The latter three facts violate hypothesis {\em (d)}.
  \hfill $\qed$
\end{proof}

Theorem \ref{uniqueness:teo} says that entailment is
the only relation
which
$(a)$ is stronger or equal than entailment,
$(b)$ extends to partial assignments the standard notion of
total-assignment satisfaction,
$(c)$ maintains the standard semantics of
$\wedge$, and
$(d)$ verifies
Property~\ref{prop:partial-entailment}(ii).
Thus, any such relation $\genericentails$ which is strictly stronger
that $\entails$ suffers for
the same ``embarrassing fact'' of
Property~\ref{prop:partial-verification}(ii).

  \begin{example}
%
    Let ``$\genericentails$'' denote a variant of $\verifies$ such
    that $\mu\genericentails\vi$ 
    implements a syntactic check that adds 
    ``$l\vee \neg l \Rightarrow \top$'', $l$ being a literal, to the
    rules used in \cref{fig:boolprop} for computing the residual
    \applymuvi{}.
    Thus, considering the formula $\vi_1$ of \cref{ex:embarassing} we
    have $\mu\genericentails\vi_1$. 

    Nevertheless,    let
$\mu\defas\set{A_1,..,A_M}$ s.t.
$\vi\defas\bigvee_{i=1}^M (A_i\wedge cube_i)$ s.t. each $cube_i$ is a
non-unary cube and $\bigvee_{i=1}^M cube_i$ is
valid and does not contain occurrences of the atoms $A_1,..,A_M$. Then 
$\mu\entails\vi$ but $\mu\not\genericentails\vi$, because
\applymuvi{} is the valid formula $\bigvee_i
cube_i$ which is not simplified by the above reduction.
If $\vi_2\defas\bigwedge\mu$, then $\vi\equiv \vi_2$,
$\mu\not\genericentails\vi$ but $\mu\genericentails\vi_2$,
whereas $\mu\entails\vi$ and $\mu\entails\vi_2$. Thus the ``embarrassing
fact'' above applies also to $\genericentails$. 
\hfill $\diamond$
\end{example}  

\ignore{
\begin{IGNOREINSHORT}
  \begin{example}
Let
$\mu\defas\set{A_1,..,A_M}$ s.t. $M<N$ and
$\vi\defas\bigvee_i (A_i\wedge cube_i)$ s.t. each $cube_i$ is a cube and $\bigvee_i cube_i$ is
valid and does not contain occurrences of the atoms $A_1,..,A_M$. Then 
$\mu\entails\vi$ but \applymuvi{} is the valid formula $\bigvee_i
cube_i$, so that $\mu\not\verifies\vi$. \hfill $\diamond$
\end{example}  
\end{IGNOREINSHORT}
}

\section{Practical consequences of using \verbverification{} or entailment}
\label{sec:practical}

In this section we analyze the practical consequences of adopting
\verbverification{} or entailment as a partial-assignment
satisfiability tests in search procedures for satisfiability and
enumeration and in formula compilation. In particular,
when entailment and verification do not coincide, we analyze the
tradeoff between the extra cost of entailment versus the benefits it may
produce. 

\subsection{\Verbverification{} vs. entailment in solving, enumeration
and compilation}
\paragraph{\Verbverification{} vs. entailment for native CNF formulas.}
  When the input formula is natively in CNF, the 
  distinction between \verbverification{} and entailment is not relevant, because these two concepts
  coincide. (We conjecture that this is the main reason why the distinction between
\verbverification{} and entailment has been long overlooked in the literature.)
For satisfiability of CNF formulas, CDCL SAT solvers directly produce
  {\em total} truth assignments, because it is not worth to perform
  intermediate satisfiability checks. 

\paragraph{\Verbverification{} vs. entailment in satisfiability.}
In  satisfiability we need
finding only one total assignment $\eta$ extending $\mu$ which satisfy
$\vi$.
Therefore
entailment produces no benefits
wrt. \verbverification{} and is more expensive, so that the latter is
always preferred.

\smallskip
The scenario changes when we deal with
enumeration-based algorithms applied to non-CNF formulas, or to
existentially-quantified formulas, or to CNF-ized formulas.

\paragraph{\Verbverification{} vs.  entailment for enumeration on
  non-CNF formulas.}
The analysis in \sref{sec:partialsat-plain} shows the following facts.
On the one hand, the advantage of adopting \verbverification{} rather than entailment for checking
partial-assignment satisfiability is that
it matches the intuition
and practical need that satisfiability checking should be fast,
since
 checking \verbverification{}
is polynomial (Property~\ref{prop:partial-verification}(iv)) and easier
to implement, whereas checking
entailment is co-NP-complete  (Property~\ref{prop:partial-entailment}(iv)),
because it is equivalent to checking the validity of the residual \applymuvi{}.
%
We stress the fact, however, that if $\mu\entails\vi$, then the residual
\applymuvi{} is typically much smaller than \vi with a much
smaller search space, since its
variables are only (a subset of) the variables which are not assigned
by $\mu$, 
so that in many actual situations the above fact could not be a major issue.

On the other hand, a main advantage of adopting entailment on
enumeration is that,
due to \cref{prop:stronger}, every
 partial assignments entailing the input formula is  a \emph{sub-assignment}
 of some other(s) \verbverifying{} it.
 Consequently, 
as soon as one verification-based enumeration procedure produces (a branch corresponding to) an assignment
$\mu$ s.t. 
$\mu\entails\vi$ but $\mu\not\verifies\vi$, it cannot realize this
fact and thus
proceeds the search to extend $\mu$ to some $\mu'\supset\mu$ s.t. 
$\mu'\verifies\vi$. This causes an extension of the search tree of up to
$2^{|\mu'|-|\mu|}$ branches.  
Therefore, for an assignment-enumeration algorithm,
 being able to enumerate partial 
assignments entailing the input formula rather than simply \verbverifying{}
it  may (even drastically) reduce the number of the satisfying assignments
enumerated. 

Also, the analysis in \sref{sec:partialsat-others} shows that
alternative forms of partial-assignment satisfiability extending
\verbverification{} ---e.g.,
adding further simplifications for the residual $\applymuvi$  like unit-propagation
or the removal of basic valid subformulas--- cannot fill 
the  gap wrt. entailment, whose theoretical properties are specific.

\paragraph{\Verbverification{} vs. entailment for
  projected enumeration.}
The analysis in \sref{sec:partialsat-exist} shows that
the above considerations for non-CNF formulas extend automatically to procedures for projected
enumeration (e.g. projected AllSAT/AllSMT, projected \#SAT/\#SMT),
even on CNF formulas, because projected enumeration
consists in enumeration on an
existentially-quantified formula $\exists \allb.\psi$, $\allb$ being
the  non-relevant atoms to be projected away.
The same considerations apply  also to other applications like
preimage computation
in symbolic model checking \cite{burch1} or
 predicate abstraction in SW verification \cite{graf_predabs97,lahiriSMTTechniquesFast2006}. 


\paragraph{\Verbverification{} vs. entailment for
  CNF-ized non-CNF formulas.}
The analysis in \sref{sec:partialsat-cnf} shows that,
although verification and entailment coincide for CNF formulas, 
the problem with non-CNF formulas cannot be fixed by simply CNF-izing a formula
upfront \cite{tseitin1,plaistedStructurepreservingClauseForm1986}
and running a CNF enumeration procedure based on partial-assignment
verification, projecting out the fresh variables introduced by the CNF-ization.
In fact, a partial assignment over the original
    atoms may entail the existentially-quantified CNF-ized
    formula without \verbverifying{} it.
Also, even adopting the  CNF-ization technique \PlaistedCNF(\NNF{\vi}) we proposed in
\cite{masinaCNFConversionDisjoint2023}, although improving
the enumeration process, does not fill the gap between
\verbverification{} and entailment in this case.

\ignore{CNF-izing upfront the non-CNF formula with the standard
  techniques \cite{tseitin1,plaistedStructurepreservingClauseForm1986}
  does not fix issues $(a)$-$(c)$, since fresh atoms are
  introduced, and \emph{a partial assignment over the original
    atoms may entail the existentially-quantified CNF-ized
    formula without \verbverifying{} it.}
  }
%

\paragraph{\Verbverification{} and entailment in formula compilation.}
With formula compilers a formula $\vi$ is encoded into a
format which makes assignment enumeration straightforward \cite{darwicheKnowledgeCompilationMap2002}.\\
d-DNNF formulas are NNF DAGs which are {\em decomposable} (i.e., all
conjunctive subformulas $\vi_1\wedge \vi_2$ are such that $\vi_1$ and
$\vi_2$ do not share atoms) and {\em deterministic} (i.e., all
disjunctive subformulas $\vi_1\vee \vi_2$ are such that $\vi_1$ and
$\vi_2$ are mutually inconsistent)  \cite{darwicheCompilerDeterministicDecomposable2002}.
The d-DNNF compilers typically implement a top-down search, implicitly
adopting \verbverification{} ($\applymuvi=\top$) 
as partial-assignment satisfiability test \cite{DarwicheDNNF-ACM01}.

OBDDs~\cite{bryant2} and SDDs~\cite{darwicheSDDNewCanonical2011} are subcases of
d-DNNFs \cite{darwicheKnowledgeCompilationMap2002} which are canonical under some
order condition  ---i.e., two equivalent subformulas \vione{} and
\vitwo{} are encoded into the same OBDD or SDD, and as such are shared
inside the DAG representation.
The OBDD and SDD compilers typically build the encoding
bottom-up, and they are able to encode 
$\applymuvi$ into $\top$ as soon as $\applymuvi$ becomes valid.
(That is, they implicitly use entailment as partial-assignment satisfiability test.)


\paragraph{\Verbverification{} and entailment for \#SMT and WMI.}
With some enumeration-based techniques (e.g. \#SMT \cite{chisticov15,PhanM15} and WMI
\cite{morettin_aij19,spallittaEnhancingSMTbasedWeighted2024a})
the total cost of {\em processing} each enumerated partial assignment dominates that of
the enumeration itself.
(E.g., with WMI each satisfying assignment $\mu$ corresponds to a polytope, and
for each $\mu$ it is necessary to compute
the integral of some multivariate function over such polytope.)
Therefore, in this context the {\em effectiveness} of enumeration in
reducing the number of assignments generated is even more important
that the {\em efficiency} in terms of CPU time required to enumerate them.   
\ignore{(Specialized techniques to reduce the number of enumerated assignments in WMI have been
proposed in \cite{morettin_aij19,spallittaEnhancingSMTbasedWeighted2024a}.)}
Therefore,  with entailment-based enumeration  the reduced number of assignments produced, and hence of
integrals computed, should definitely compensate the extra cost of
entailment checking.

\subsection{\Verbverification{} vs. entailment in search
  procedures and formula compilers.}
When applied to satisfiable formulas,
algorithms like \new{Analytic Tableaux}~\footnote{Notice that Analytic Tableaux may generate duplicated
 or subsumed  assignments (see \cite{dagostino1,gs-infocomp2000})}
\cite{smullyan1} 
or ``classic'' \new{DPLL}~%
\footnote{Classic DPLL procedure \cite{davis7} was
  designed to work for CNF  
  formulas. Nevertheless it is easy to produce a non-CNF version of this
procedure (see e.g. \cite{armando5}).} 
\cite{davis7}, or even recent sophisticated AllSAT procedures for circuits~\cite{friedAllSATCombinationalCircuits2023},
 produce branches representing
partial assignments which {\em \verbverify} the input formula.
The same happens with enumeration algorithms for d-DNNF formulas,
which resemble the  tableaux enumeration algorithm,
with the restriction that the formulas they are applied to
are deterministic and decomposable NNFs, 
so that
the enumerated assignment are pairwise disjoint and straighforward to
compute, because the enumeration reduces to the traversal of the and-or 
DAG without encountering inconsistent branches
\cite{DarwicheDNNF-ACM01}.

\ignore{
  Consequently, 
as soon as these procedures produce (a branch corresponding to) an assignment
$\mu$ s.t. 
$\mu\entails\vi$ but $\mu\not\verifies\vi$, they cannot realize this
fact and
proceed the search until they extend it to some $\mu'\supset\mu$ s.t. 
$\mu'\verifies\vi$. This causes an extension of the search tree of up to
$2^{|\mu'|-|\mu|}$ branches.  
}

\begin{figure}[t]
  \begin{tabular}{lll}
    \scalebox{.8}{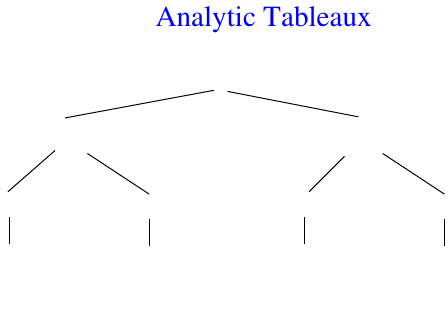} 
    &\ \hspace{.5cm} &
    \scalebox{.8}{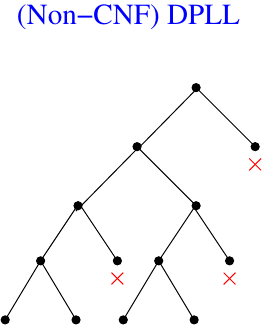}       
  \end{tabular}
  \caption{\label{fig:search} $\vi\defas\mainwff{}$.
    \newline
    Enumeration by  analytic tableaux (left) and (non-CNF) DPLL (right).
}
\ \\
  \begin{tabular}{lllll}
    \scalebox{.8}{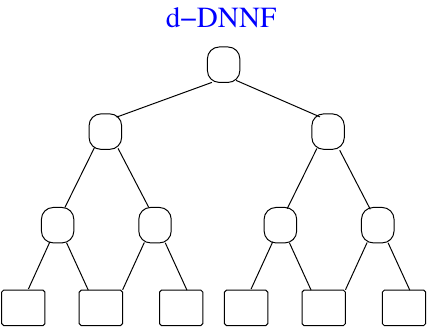} 
    &\ \hspace{.2cm} &
    \scalebox{.8}{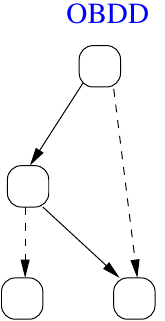} 
    &\ \hspace{.2cm} &
    \scalebox{.8}{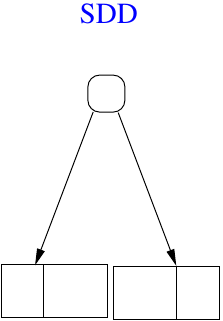} 
  \end{tabular}
  \caption{\label{fig:compilers} $\vi\defas\mainwff{}$.
  \newline Compilation of $\vi$ into: d-DNNF (left), OBDD (center), SDD (right).
\\ }  
%
\end{figure}

\begin{example}
\label{ex:enumeration-basic}
Consider $\vi\defas\mainwff$.\\
Notice that the single assignment $\mu\defas\singlemu{}$
subsumes all total truth assignments satisfying \vi and
is such that
$\mu\entails\vi$ but $\mu\not
\verifies\vi$, since $\applymuvi=(\Atwo\vee\nAtwo)\wedge(\Afour\vee\nAfour)$.

Figure~\ref{fig:search} 
represents 
the search trees corresponding to 
AllSAT executions of Analytic
Tableaux and 
(non-CNF) DPLL  
on $\vi$%
\footnote{Here in DPLL the pure-literal rule \cite{davis7} is not used because 
 in All-SAT it may hinder the enumeration of
 relevant models (see, e.g., \cite{sebastiani_frocos07}).}
\footnote{The search trees may vary with the ordering by
  which variables and subformulas are analyzed.}.
Since (the DAG version of) $\vi$ is already in d-DNNF
(\cref{fig:compilers}), in this case the d-DNNF enumeration resembles the
tableaux enumeration. 
In all cases, the enumeration produces:~%
\\\allmus{}.
Notice that all the assignments produced are total.
Notice also that neither Analytic Tableaux nor DPLL nor d-DNNF
enumeration in this case can produce
\singlemuset{}, regardless the adopted strategy.
The same applies also to the \verbverification-based AllSAT circuit
enumeration of \cite{friedAllSATCombinationalCircuits2023}, see \cref{ex:enumeration-entail}.
\hfill$\diamond$
\end{example}

OBDDs \cite{bryant2} contain branches representing
partial assignments which {\em entail} the input formula, because if $\mu\entails\vi$ then
\applymuvi is
valid (Property~\ref{prop:partial-entailment}(iii)), so that its corresponding 
sub-OBDD is reduced into the $\top$ node. (SDDs~\cite{darwicheSDDNewCanonical2011} behave similarly.)

\begin{example}
\label{ex:compilation}
Consider $\vi$ as in \cref{ex:enumeration-basic}.
  Figure~\ref{fig:compilers} represents possible compilations of $\vi$
  into d-DNNF, OBDD and SDD respectively.~\footnote{The outcome may
    vary with the ordering by
  which variables and subformulas are analyzed.}
Notice that, unlike with OBDDs and SDDs which collapse equivalent
subformulas into one,  the two $\vee$-subformulas in the d-DNNF are not
recognized to be equivalent to the nodes $\Aone$ and $\nAthree$
respectively. 

Applying enumeration to the d-DNNF we obtain the total
assignments of \cref{ex:enumeration-basic}.
By applying enumeration to the OBDD or to the SDD, we obtain
the single-assignment \singlemuset{}, so that $\singlemu\entails\vi$ but $\singlemu\not
\verifies\vi$.
\hfill$\diamond$
\end{example}

\subsection{Implementing entailment within enumeration procedures}
%
Implementing efficiently entailment-based enumeration procedures can be
tricky. One possible approach is to
integrate AllSAT/AllSMT with OBDDs, exploiting the capability of OBDD
to implicitly perform entailment-based reductions \cite{cavada_fmcad07_predabs}. The main problem
with this approach is the fact that OBDDs quickly blow up in memory as
soon as the input formula increases in size.

\cite{FazekasSB16,mohleDualizingProjectedModel2018} proposed a formal
reasoning framework based on the idea of {\em dualized search}:
while the main enumerator produces partial assignments $\mu$ for the input
formula $\vi$, another ``dual'' SAT solver is incrementally called on
$\TseitinCNF(\neg\vi)\wedge\andmu$: if it is unsatisfiable
then the main enumerator produces  $\mu$ and
backtracks. 
Based on the analysis in \cite{sebastianiAreYouSatisfied2020} and
noticing that the above dual check in
\cite{FazekasSB16,mohleDualizingProjectedModel2018} is an entailment
($\mu\entails\vi$) rather than a \verbverification{} ($\mu\verifies\vi$), in \cite{moehle20,moehle21} we
proposed enumeration and model counting frameworks and algorithms
based on dual search, enhanced with chronological backtracking. 
Unfortunately no
implementation efficient enough to compete with state-of-the-art enumeration
procedures was produced.

\cite{friedAllSATCombinationalCircuits2023} presented {\sf HALL}, a sophisticated AllSAT
procedure for circuits which, among other techniques, exploited a form of
(\verbverification-based) {\em generalization} (aka minimization or reduction): as soon as a 
total truth assignment $\eta$ satisfying $\vi$ is produced,
the literals in $\eta$ are removed one by one and the resulting
partial assignment is tested to \verbverify{} $\vi$.
In \cite{friedEntailingGeneralizationBoosts2024}
we
 have enhanced {\sf HALL} generalization procedure by substituting
 \verbverification{} checks with entailment checks based on dualized
 search, boosting its performance in terms of both time efficiency and size of 
 partial-assignment sets.

\begin{example}
\label{ex:enumeration-entail}
Consider $\vi\defas\mainwff$\\ as in \cref{ex:enumeration-basic} and
assume some total assignment satisfying it is produced,
e.g. $\eta\defas \set{\Aone,\Atwo,\nAthree,\Afour}$.
Since there is no literal $l\in\eta$
s.t. $\eta\backslash\set{l}\verifies\vi$, no assignment
reduction based on verification can be successfully applied to $\eta$.
Consequently, even the sophisticated \verbverification{}-based circuit AllSAT procedure in
\cite{friedAllSATCombinationalCircuits2023} produces the set of four
un-reduced total assignments of \cref{ex:enumeration-basic}.

Instead, since $\eta\backslash\set{\Atwo,\Afour}\entails\vi$
whereas $\eta\backslash\set{\Aone}\not\entails\vi$ and
$\eta\backslash\set{\nAthree}\not\entails\vi$,
the entailment-based assignment-reduction technique of
\cite{friedEntailingGeneralizationBoosts2024}  shrinks $\eta$ into
$\singlemu$, so that the enumeration procedure returns $\singlemuset$.
\hfill$\diamond$
\end{example}

\ignore{
\TODO{Aggiunte da prima, piazzare da qualche parte:\\}
In many application domains, fundamental operations ---like {\em 
\RSTODO{rivedere questo paragrafo?}
  preimage computation} in symbolic model checking
(see e.g. \cite{burch1}) or {\em predicate abstraction} in SW verification
(see e.g. \cite{graf_predabs97,beyercgks09})---
require dealing with existentially-quantified formulas and with
the enumeration of partial assignments satisfying them.
(This applies implicitly also to a formula $\psi$ resulting from
Tseitin CNF-ization of some formula $\vi$, because
$\vi(\alla)\equiv\exists.\allb\psi(\alla,\allb)$ (see \sref{sec:background}).)
} 

\ignore{
\subsection{A Relevant Example Application: Predicate Abstraction.}
\label{sec:partialsat-predabs}
 
Given a propositional formula $\phi$ on $\allb$ and a set
$\mathbf{\Phi}\defas\set{\phi_i}_{i}$ of formulas on $\allb$ denoting relevant
``predicates'' and a set $\alla$ of fresh atom s.t. each $A_i$
labels $\phi_i$, 
then the {\em Predicate Abstraction} of $\phi$ wrt. $\mathbf{\Phi}$ is
defined as follows \cite{graf_predabs97}:
\begin{eqnarray}
  \label{eq:predabs}
 \predabsphi \defas
\exists \allb. 
( \phi \wedge \bigwedge_{i} (A_i\iff \phi_i)
) .
\end{eqnarray}
 \predabsphi{}
is typically computed as {\em projected anumeration}, i.e., as a disjunction of mutually-inconsistent partial
assignments (cubes) $\mu_j$ on \alla s.t. $\mu_j\entails
\predabsphi$ and 
$\bigvee_j\mu_j$ is equivalent to \predabsphi
\cite{lahiriSMTTechniquesFast2006,cavada_fmcad07_predabs}.
\footnote{Notice that predicate Abstraction is most often referred to SMT formulas
  $\phi$ and $\set{\phi_i}_i$, so that \eqref{eq:predabs} involves
  also the existential quantification of first-order theory-specific variables
  and $\mu_i$ are theory-consistent SMT assignments
  \cite{graf_predabs97,lahiriSMTTechniquesFast2006,cavada_fmcad07_predabs}.
However, restricting our discussion to the purely-propositional case suffices for our
purposes and makes the explanation much simpler.}

We notice that in the computation of such cubes the distinction between $\verifies$ and
$\entails$ may be very relevant: whereas 
it would be desirable to look for partial assignments $\mu_j$ 
{\em entailing} \predabsphi{}  to keep them small and hence reduce
their number,
 most algorithms can reveal
only when $\mu_j$ 
{\em \verbverifies{}} it, and 
are thus
incapable of producing partial assignments 
$\mu_j$ s.t. $\mu_j\entails\predabsphi$ and
$\mu_j\not\verifies\predabsphi$.
This happens every time that, 
for some $A_k$ and some $\mu^k$
on (subsets  of) $\alla\setminus \set{A_k}$,
{\em both} 
$\apply{(\phi\wedge\bigwedge_{i\neq k}(A_i\iff \phi_i))}{\mu^k} \wedge \phi_k$
and 
$\apply{(\phi\wedge\bigwedge_{i\neq k}(A_i\iff \phi_i))}{\mu^k} \wedge
\neg \phi_k$
 are satisfiable but they are satisfied by {\em distinct} sets of assignmets
 $\delta$ on \allb (Definition~\ref{def:entailment-exist}), so that
 $\mu_k\entails\predabsphi$ but 
 $\mu_k\not\verifies\predabsphi$.

\begin{example}
Consider the CNF formula
$\phi\defas(\neg B_1\vee B_2)\wedge (B_1\vee \neg B_2)$
and the ''predicate'' CNF formulas
$\Phi_1\defas B_1\wedge B_2$ and
$\Phi_2\defas \neg B_1\wedge B_2$.
Then
\begin{eqnarray}
  \label{eq:predabs1}
  \predabsphi &\defas & \exists B_1 B_2. \left (
 \begin{array}{lll}
(\neg B_1\vee B_2)\wedge (B_1\vee \neg B_2) & \wedge \\
(A_1 \iff (B_1\wedge B_2)) & \wedge \\
(A_2 \iff (B_1\wedge \neg B_2))
 \end{array}
  \right )
  \\
  \label{eq:predabs2}
  &\Leftrightarrow& (A_1 \wedge \neg A_2) \vee (\neg A_1 \wedge \neg A_2) 
  \\
  \label{eq:predabs3}
  &\Leftrightarrow& \neg A_2
\end{eqnarray}
Both \set{A_1,\neg A_2} and \set{\neg A_1,\neg A_2}
\verbverify{} \predabsphi,
whereas  \set{A_1} entails it without \verbverifying{} it.
Thus, if the algorithm is able to detect if
$\mu_j\entails\predabsphi$ and
$\mu_j\not\verifies\predabsphi$, then it can return \eqref{eq:predabs3},
otherwise it can only return \eqref{eq:predabs2}.
\hfill $\diamond$
\end{example}

%
\noindent
Therefore, having algorithms able to stop extending $\mu_j$ as soon as 
$\mu_j\entails\predabsphi$,  even when 
$\mu_j\not\verifies\predabsphi$,  would produce much more compact
 formulas. 
}

\ignore{
\section{A simple poll}
\label{sec:poll}

\pagestyle{plain}
\pagenumbering{roman}

\pagestyle{plain}
\pagenumbering{arabic}

\title{%
  A Poll on the Meaning of Partial-Assignment Satisfiability
%
%
}

\author{
Roberto Sebastiani 
}

\institute{%
DISI, University of Trento, Italy%
}

\date{\today}

\maketitle
\ignoreinshort{
\large
\begin{center}
\noi
{\em Latest update: \today}
\end{center}
}

\begin{center}
\today
\end{center}


\label{sec:}

\section*{Premise}
This document reports the text and results of a poll I made among 65
members of current and past PC of SAT and SMT conference on the notion
of ``partial-assignment satisfiability''. Unfortunately only 20 people
replied. I asked each person to say
whether the following statements  were true or false according his/her
intended meaning of the sentence  ``a partial truth assignment
satisfies...''.

\section*{Statements}

\begin{enumerate}

\item
A {\em partial} truth assignment $\mu$ on $\alla$ satisfies $\vi$
if and only if all the total truth assignments $\eta$ on $\alla$ 
which extend $\mu$ satisfy $\vi$.

\item
A {\em partial} truth assignment $\mu$ on $\alla$ satisfies $\vi$
if and only if the result of applying $\mu$ to $\vi$ is $\top$.

\item
Let $\vi$ and $\vione$ be logically equivalent. 
Then a {\em partial} truth assignment $\mu$ on $\alla$ satisfies $\vi$
if and only if $\mu$  satisfies $\vione$.


\ignore{
\item
Checking
if a {\em partial} truth assignment $\mu$ on $\alla$ satisfies $\vi$
requires at most a polynomial amounts of steps.
}

\item
A {\em partial} truth assignment $\mu$ on $\alla$ satisfies $\exists \allb.\ps$
if and only if, there exists a total truth assignment $\gamma$ on $\allb$
s.t., for every total truth assignment $\eta$ extending $\mu$,
$\eta\cup\gamma$ satisfies $\ps$. 

\item
A {\em partial} truth assignment $\mu$ on $\alla$ satisfies $\exists
\allb.\ps$ if and only if, there exists a total truth assignment 
$\gamma$ on $\allb$ s.t. the result of applying $\mu\cup\gamma$ to 
$\ps$ is $\top$.

\end{enumerate}

\section*{Legenda}
In the above statements:
\begin{itemize}
\item 
$\alla\defas\set{A_1,...,A_N}$,
$\allb\defas\set{B_1,...,B_K}$ 
are disjoint sets of propositional atoms,
\item  
$\vi$, $\vione$  are generic propositional formulas over $\alla$, 
\item 
 $\ps$ is a generic propositional formula over $\alla\cup\allb$, 
\item 
$\mu$ 
is a {\em partial} truth assignment on $\alla$ (i.e., $\mu$ assigns a 
value only to a subset of $\alla$)
\item 
$\eta$ 
is a {\em total} truth assignment on $\alla$ (i.e., $\eta$ assigns a 
value to all atoms in $\alla$)
\item 
$\gamma$ 
is a {\em total} truth assignment on $\allb$ (i.e., $\gamma$ assigns a 
value to all atoms in $\allb$)
\end{itemize}
By ``apply a (total or partial) assignment $\mu$ to $\vi$'' we mean 
``substitute all instances of each assigned
$A_i$ in $\vi$ with the truth value in $\set{\top,\bot}$  assigned by
$\mu$ and apply the standard
propagation of truth values through the Boolean connectives
(i.e., $\neg\top \thus \bot$, 
$\neg\bot \thus \top$,
$\phi\wedge\top \thus \phi$, 
$\phi\wedge\bot \thus \bot$, and the corresponding ones for $\vee$, $\imp$, $\iff$).''

\newpage
\section*{Results}
Here follow the complete results (table \ref{tab:results} left) and
its summary, stating the number of people providing each combination
of answers (table \ref{tab:results_summary} right).  
All the 20 repliers have been pointed to this document.
The names of the repliers are kept anonymous. 
To each person who replied, however, I gave a code number so that he/she can
crosscheck his/her answer in the table. The code number is completely
uncorrelated with the alphabetic ordering of the names or surnames of
the repliers.

\begin{table}[t]
  \begin{tabular}{lll}
  \begin{tabular}{|r|l|l|l|l|l|}
\hline
Replier Code  &    \#1 & \#2 & \#3 &  \#4 & \#5\\
\hline
1 & true & false & true & true & true\\
2 & true & true & true & true & true\\
3 & false & true & false & false & true\\
4 & true & false & true & true & false\\
5 & true & false & true & true & false\\
6 & true &  & true & false & \\
7 & false & true & false & false & true\\
8 & false & true & false & false & true\\
9 & true & true & true & false & false\\
10 & false & true & false & false & true\\
11 & false & true & false & false & true\\
12 & true & true & false & true & true\\
13 & false & true & true & false & true\\
14 & true & false & true & false & false\\
15 & true & true & true & true & true\\
16 & true & true & true & false & false\\
17 & true & false & true & true & false\\
18 & true & false & true & true & false\\
19 & false & true & true & false & true\\
20 & true & true &  & true & true\\
\hline
  \end{tabular}
  \vspace{.5cm}
& & 
  \centering
  \begin{tabular}{|l|l|l|l|l|r|}
    \hline
    \#1 & \#2 & \#3 &  \#4 & \#5 & \# of repliers\\
    \hline
false & true & false & false & true & 5\\
false & true & true & false & true & 2\\
\hline
true & false & true & false & false & 1\\
true &  & true & false &  & 1\\
true & false & true & true & false & 4\\
true & false & true & true & true & 1\\
\hline
true & true & false & true & true & 1\\
true & true & false & true & true & 2\\
true & true & true & true & true & 2\\
 true & true &  & true & true & 1 \\
\hline
 &  &  &  &  & 20\\
\hline
  \end{tabular}
  \end{tabular}
  \caption{Left: Complete results of the poll. The code number of the left is the
    code given to each person who replied for cross-checking his/her
    reply.
  A blank is when a replier omitted giving an answer to the specific
  question.
\newline Right: Summary of the results.}
  \label{tab:results}
  \label{tab:results_summary}
\end{table}

\section*{Comments}

Some comments are in order. The questions refer to a short
paper I wrote on the
topic  \cite{sat19}.
%
Statements 1, 2, 3, 4, 5 refer, respectively,
to Definition 1 (entailment), Definition 2 (\verbverification),
Property 3(ii), Remark 1, and Definition 4 in \cite{sat19}.

\begin{itemize}
\item \#1=true means that one sees partial-assignment
  satisfiability as entailment (Definition 1 in \cite{sat19}). 
\item \#2=true means that one sees partial-assignment
  satisfiability as \verbverification (Definition 2 in \cite{sat19}). 
\item According to Proposition 1 in \cite{sat19}, {\em for generic formulas}
  \#1=true and
  \#2=true are not compatible, the latter being strictly stronger than
  the former. (They are equivalent only for
  tautology-free CNF formulas.)
\item \#3=true holds iff \#1=true (entailment) but not iff
\#2=true (\verbverification). See properties 3(ii) and 4(ii). 
\item If we define entailment \resp{\verbverification} of $\exists\allb.\psi$ as
  the entailment \resp{\verbverification} of its Shannon Expansion, 
  then we see that \#4=true is 
  incompatible with both entailment and \verbverification, whereas \#5=true 
means that one sees partial-assignment
  satisfiability as \verbverification. (Definition 3 and 4 and
  Properties 5 and 6 in \cite{sat19})
\end{itemize}
\noindent
To this extent, we would have answered one of the two 
alternative configurations:

\begin{center}
  \begin{tabular}{|l|l|l|l|l|c|}
    \hline
    \#1 & \#2 & \#3 &  \#4 & \#5 & \\
    \hline
true & false & true & false & false & \mbox{``satisfiability is entailment''} \\
false & true & false & false & true & \mbox{``satisfiability is \verbverification''}\\
    \hline
  \end{tabular}  
\end{center}

\noindent
Of course \#4 and \#5 depend on the assumption of defining entailment \resp{\verbverification} of $\exists\allb.\psi$ as
  the entailment \resp{\verbverification} of its Shannon Expansion. 
Also, the result may be effected by the fact that maybe we did not stress
enough the fact that ``generic formulas'' included non-CNF ones.
\ \\

Overall, although only 20 out of 65 people replied, 
the result 
reveals the lack of general agreement among the repliers on the notion of
partial-assignment satisfiability and on its consequences: e.g.,
the repliers split evenly into supporters 
of entailment, of \verbverification, and people considering them equivalent.

\FloatBarrier
\bibliographystyle{abbrv} 
\bibliography{rs_refs,rs_ownrefs,rs_specific,sathandbook}

}
\section{Conclusions and Future Work}
\label{sec:concl}
In this paper we have analyzed in depth the issue of formula satisfaction by
partial assignments.
We have identified two alternative notions that are implicitly used in the literature,
namely {verification} and {entailment}, and we have showed that, 
although the former is easier to check and as such is implicitly used
by most current search procedures, the latter has better theoretical
properties, and
can improve the efficiency and effectiveness of enumeration
procedures.

From a theoretical viewpoint,
we highlight the need for a unique universally-agreed definition of
partial-assignment satisfaction
\footnote{In 2020 we ran an online poll among SAT and SMT PC members about the meaning of partial-assignment satisfiability: 35\% said ``entailment'', 35\% said ``\verbverification{}'', 30\% said both (!).}, 
and we champion the idea that it
should be defined as
entailment, 
considering  \verbverification{} an easy-to-check sufficient condition for it.
From a practical viewpoint, we suggest that entailment should be
adopted as partial-assignment satisfaction check inside
enumeration-based procedures in place of \verbverification, so that to
reduce the number of enumerated assignments.

\smallskip
The analysis presented in this paper opens several research
avenues. 
First, we wish to investigate and evaluate empirically novel entailment-based
techniques for both (projected) enumeration and (projected) model counting.
In particular, we wish to enhance our (projected) AllSAT procedure
in \cite{spallittaDisjointPartialEnumeration2024,spallittaDisjointPartialEnumerationIJA2025} with entailment
tests.
Second, we wish to implement entailment-based enumeration techniques
to enhance our AllSMT procedures in \mathsat{} \cite{mathsat5_tacas13} and evaluate them empirically. In particular,
we wish to embed and evaluate such entailment-based AllSMT procedure inside our WMI tools
\cite{morettin_aij19,spallittaEnhancingSMTbasedWeighted2024a}. 
Third, we wish to embed novel entailment-based AllSMT procedures
inside our SMT compilers for \T-OBDDs and \T-SDDs
\cite{MicheluttiEcai24}, extending it also to generic \T-d-DNNFs.

\newpage
\pagenumbering{roman}
\FloatBarrier
\section*{Acknowledgements} 

The analysis described in this paper has
strongly benefitted from interesting discussions, either in person or via
email, with (in alphabetic order):
Armin Biere,
Alessandro Cimatti,
Dror Fried,
Allen van Gelder (R.I.P.),
Alberto Griggio,
Gabriele Masina,
David Mitchell,
Sibylle M\"ohle,
Paolo Morettin,
Andrea Passerini,
Alexander Nadel,
Yogev Shalmon,
Laurent Simon,
Giuseppe Spallitta,
Armando Tacchella,
Stefano Tonetta,
and
Moshe Vardi,
whom are all warmly thanked.


The author was supported in part by the MUR PNRR project FAIR - Future AI
Research (PE00000013) funded by the NextGenerationEU;
he was partially funded under the NRRP, Mission
4 Component 2 Investment 1.4, by the European Union – NextGenerationEU (proj. nr. CN 00000013);
he was supported in part by the TANGO project funded by the EU Horizon Europe research and innovation program
under GA No 101120763. 
Views and opinions expressed are however those of
the author only and do not necessarily reflect those of the European Union, the European Health and Digital
Executive Agency (HaDEA) or The European Research Council. Neither the European Union nor the granting
authority can be held responsible for them.

\bibliographystyle{abbrv} 
\bibliography{rs_refs,rs_ownrefs,rs_specific,sathandbook}


\end{document}